\documentclass[12pt,preprint]{aastex}

\slugcomment{Accepted to ApJ}

\def\MO{M_\odot}

\def\CO{\mathrm{C}^{18}\mathrm{O}}
\def\HCO{\mathrm{H}^{13}\mathrm{CO}^{+}}

\def\MCLU{M_{\mathrm{clump}}}

\def\dv{\Delta{V}}
\def\dvC{\Delta{V}_{\mathrm{clump}}}
\def\RC{R_{\mathrm{clump}}}

\shorttitle{Survey of Cluster-Forming Clumps in $\HCO$(1--0) line emission}
\shortauthors{A.E.Higuchi et al.}

\begin{document}

\title{A Mapping Survey of Dense Clumps Associated with Embedded Clusters II : 
Can Clump-Clump Collisions Induce Stellar Clusters?}

\author{Aya E. HIGUCHI\altaffilmark{1}, Yasutaka KURONO\altaffilmark{1}, Masao SAITO\altaffilmark{1}, 
\and  Ryohei KAWABE\altaffilmark{2}}
\email{aya.higuchi@nao.ac.jp}

\altaffiltext{1}{National Astronomical Observatory of Japan 2-21-1 Osawa, Mitaka, Tokyo, 181-8588, Japan}
\altaffiltext{2}{Nobeyama Radio Observatory, Nobeyama, Minamimaki, Minamisaku, Nagano 384-1305, Japan}

\begin{abstract}

We report the $\HCO$ ($J$=1--0) survey observations toward embedded clusters obtained using the Nobeyama 45m telescope, 
which were performed to follow up our previous study in the $\CO$ survey with a dense gas tracer.
Our aim is to address the evolution of cluster-forming clumps.
We observed the same 14 clusters in $\CO$, which are located at distances from 0.3 to 2.1$\,$kpc
with 27$''$ resolution (corresponding to Jeans length for most of our targets) in $\HCO$.
We detected the 13 clumps in $\HCO$ line emission and obtained the physical parameters of the clumps with radii of 0.24--0.75$\,$pc, 
masses of 100--1400$\,$$\MO$, and velocity widths in FWHM of 1.5--4.0$\,$${\rm{km \: s}^{-1}}$.
The mean density is $\sim$ 3.9 $\times$ 10$^{4}$ cm$^{-3}$ and the equivalent Jeans length is $\sim$ 0.13$\,$pc at 20$\,$K.
We classified the $\HCO$ clumps into three types, $\it{Type \ A}$, $\it{Type \ B}$, and $\it{Type \ C}$ according to the relative locations of the $\HCO$ clumps and the clusters (see our previous study).
Our classification represents an evolutionary trend of cluster-forming clumps because dense clumps 
are expected to be converted into stellar constituents, or dispersed by stellar activities.
We found a similar but clearer trend than our previous results for derived star formation efficiencies to increase 
from $\it{Type \ A}$ to $C$ in the $\HCO$ data, and for the dense gas regions within the clumps traced by $\HCO$ to be 
sensitive to the physical evolution of clump-cluster systems. 
In addition, we found that four out of 13 $\HCO$ clumps which we named ``DVSOs" (Distinct Velocity Structure Objects) 
have distinct velocity gradients at the central parts of them, i.e., at the location of the embedded clusters.
Assuming that the velocity gradients represent the rigid-like rotation of the clumps, 
we calculated the virial parameter of the $\HCO$ clumps by taking into account the contribution of rotation, and
found that the DVSOs tend to be gravitationally unbound. 
In order to explain the above physical properties for DVSOs in a consistent way, 
we propose a clump-clump collision model as a possible mechanism for triggering formation of clusters.

\end{abstract}
\keywords{ISM:clouds --- radio lines:stars--- stars:formation }

\section{INTRODUCTION}

Almost all stars (more than 90$\,$$\%$ of stars within our galaxy), in particular massive stars ($>$$\,$8$\,$$\MO$) 
within the disk of the Milky Way form as members of clusters \citep{lad03}. 
Thus, clusters have been long considered as important laboratories for astronomy.
The mechanisms of cluster formation triggered by external effects have been proposed from both observations and theoretical studies.
For example, the effects from H {\sc ii} regions  \citep[e.g.,][]{hes05}, supernovae  \citep[e.g.,][]{yas06}, and molecular outflows
\citep[e.g.,][]{shi08} toward the cluster forming regions have been well studied previously.
In addition, Furukawa et al. (2009) argued that cluster formation can be triggered by the dynamical interaction of massive molecular clouds. 
However, evidence for triggered cluster formation via cloud-cloud interactions is limited to morphological evidence for most cases.

In recent studies, the cluster-forming clumps, which are the dense regions in molecular clouds 
(size $\sim$ 1 pc, mass $\sim$ 100--1000$\,$$\MO$, density $\sim$ 10$^{3-5}$$\,$$\rm{cm}^{-3}$) 
are considered to be the parental objects of clusters \citep{lad03}.
Ridge et al. (2003) have carried out a survey of the dense clumps associated with the embedded clusters using the 
$^{13}\rm{CO}$($J$=1--0), $\CO$($J$=1--0), and $\CO$($J$=2--1) molecular lines.
They proposed a scheme for classifying the clumps and revealed that the peak positions of the clumps in the evolved stages 
are apart from the Herbig Ae/Be stars or $\it{IRAS}$ sources. 
However, the spatial resolution of their observations was insufficient to resolve the components whose size is comparable to the Jeans length. 
Higuchi et al. (2009) carried out a survey of the dense clumps associated with 14 embedded clusters in high resolution $\CO$ ($J$=1--0) observations with the Nobeyama 45m telescope. 
From the spatial relation between the clump structures and forming clusters, 
they classified the $\CO$ clumps into three types ($\it{Type \ A, \ B, \rm{and}}$ $\it{C}$).
They revealed the evolutionary sequence from $\it{Type \ A}$ to $\it{C}$ with dispersing gaseous components of the clumps.
Moreover, they found that the star formation efficiencies (SFEs) of the dense clumps have a tendency to increase from 
$\it{Type \ A}$ to $\it{C}$, which supports the notion that their classification shows the evolutionary stages of the cluster-forming clumps.

In this paper, we present high-resolution observations with the Nobeyama 45m telescope toward the 14 clusters.
We used $\HCO$($J$=1--0) molecular line emission whose critical density ($\sim$ 10$^{5}$$\,$$\rm{cm}^{-3}$)
is an order of magnitude larger than that of the $\CO$ line.
Ikeda et al. (2007) found a good correlation of the spatial distribution of the dust continuum emission with that of the
$\HCO$($J=$1--0) line emission and identified star-forming dense cores (size $\sim$ 0.1 pc, mass $\sim$ 10$\,$$\MO$, 
density $\sim$ 10$^{5}$$\,$$\rm{cm}^{-3}$) with the $\HCO$($J$=1--0) line.
The $\HCO$ line emission traces the denser regions within the clumps associated with the embedded clusters. 
Out of our 14 targets, six are common with the sample in the study of Ridge et al. (2003), which are located within only 1$\,$kpc. 
We morphologically classify the cluster-forming clumps according to the spatial relation between the 
clump structures and the location of clusters.
We also discuss the evolution of the dense clumps and the formation mechanism of the clusters.

\section{OBSERVATIONS}

We have carried out $\HCO$($J$=1--0; 86.754330 GHz) observations using the 45m telescope at the Nobeyama Radio Observatory 
(NRO) from December 2005 to May 2008 toward the same targets as Higuchi et al. (2009).
We used the 25-BEam Array Receiver System (BEARS) for the front end receiver, which had 5 $\times$ 5 beams separated 
by 41$^{\prime\prime}$.1 on the plane of the sky \citep{sun00, yam00}.
At 87 GHz, the main beam size was 18$''$ in FWHM and the main beam efficiency ($\eta$) was 0.51.
At the back end, we used 25 sets of 1024 channel auto-correlators (AC), which had 32 MHz bandwidth and 37.8 kHz 
frequency resolution \citep{sor00}. 
The frequency resolution corresponds to the velocity resolution of 0.13$\,$km $\rm{s}^{-1}$ at 87 GHz.
Our mapping observations were made using the on-the-fly mapping technique \citep{saw08} and covered approximately 5$^{\prime}$ $\times$ 5$^{\prime}$--10$^{\prime} \times 10^{\prime}$ areas which corresponds to 3.2$\,$pc $\times$ 3.2$\,$pc at a distance of 2.1$\,$kpc centered on the position of the clusters. 
During the observations, the double sideband system noise temperatures were in the range between 250$\,$K and 350$\,$K. 
We used the emission-free area near the observed sources as the off positions. 
The standard chopper wheel method was used to calibrate the intensity into ${T}^{*}_{\mathrm{A}}$, 
the antenna temperature corrected mainly for the atmospheric attenuation. 
The telescope pointing was checked every 1.5 hr by observing SiO maser sources near (within $\sim$ 45$'$ for all target objects) the target objects. 
The pointing accuracy was better than 3$^{\prime\prime}$.
To correct the difference in intensity scales among the 25 beams of the BEARS, 
we used calibration data obtained from the observations toward the center of S140 region using the SIS receiver equipped with 
a single sideband (SSB) filter, ``S100", and the acousto-optical spectrometer (AOS) as a backend.

We used a convolution scheme with a spheroidal function \citep{saw08} to calculate the intensity at each grid point of the final map-cube 
(R.A., decl., and $v_{\mathrm{LSR}}$) data with a spatial grid size of 9$^{\prime\prime}$, which is 
half of the beamsize; the final effective resolution is 27$^{\prime\prime}$ and the effective integration time at each 
grid point is about 160 sec. 
As a result, the mean rms noise level for all of the objects were 0.15 K in ${T}^{*}_{\mathrm{A}}$ 
with a velocity resolution of 0.13$\,$km $\rm{s}^{-1}$.

\section{RESULTS}

\subsection{Total integrated intensity and velocity dispersion maps of the $\HCO$ emission}\label{3-1}

\begin{figure}
\epsscale{0.9}
\plotone{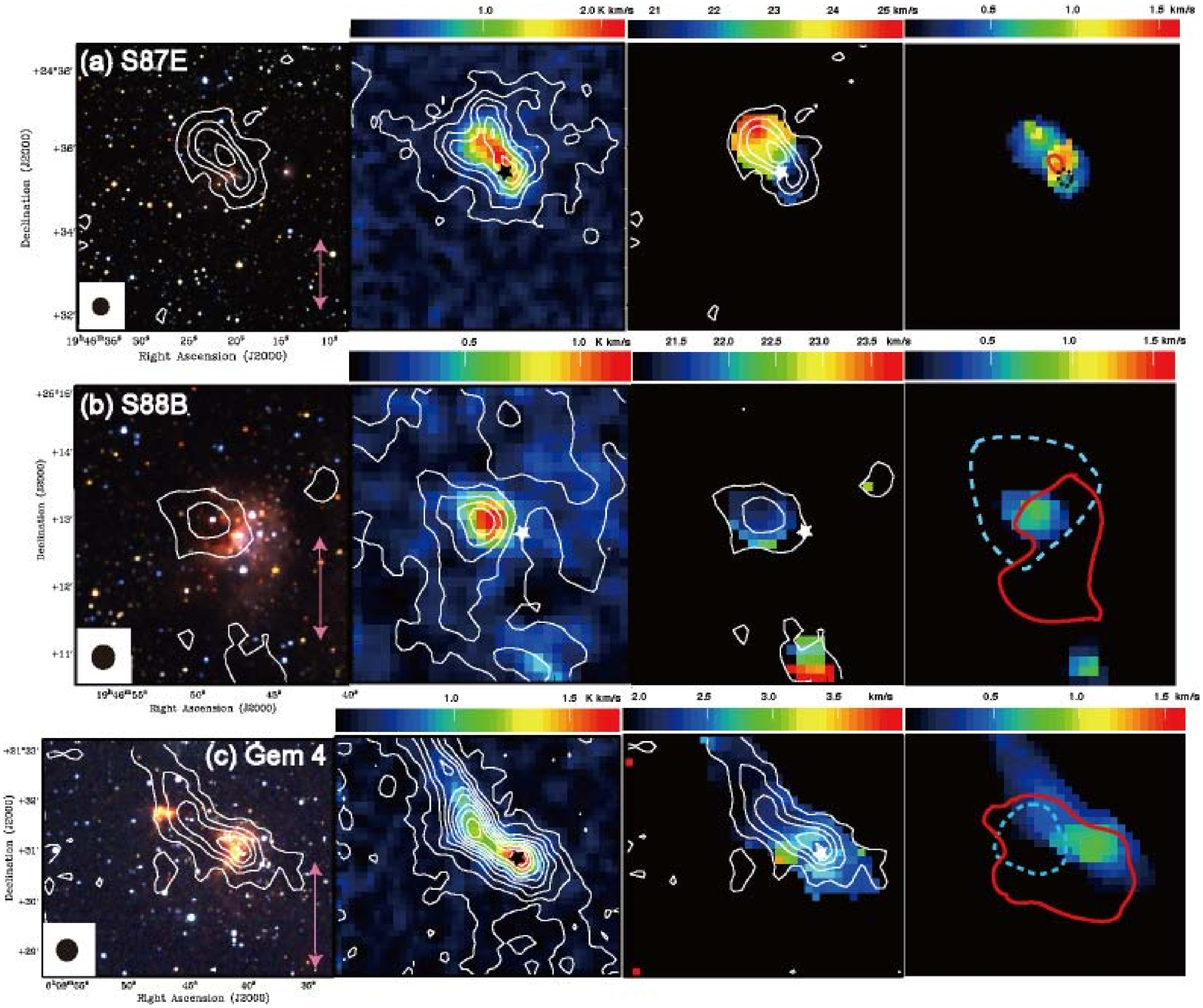}
\caption{
Total integrated intensity maps of the $\HCO$($J$=1--0) emission 
(contours) superposed on the $JHK_\mathrm{s}$ composite color images in log scale from 2MASS (left),
total integrated intensity maps of the $\CO$($J$=1--0) emission (contours; Higuchi et al. 2009) 
superposed on the $\HCO$($J$=1--0) emission  (middle-left), $\HCO$($J$=1--0) emission (contours)
superposed on the 1st moment maps (middle-right), 
and velocity dispersion maps (right) for S87E (a), S88B (b), and Gem 4 (c). 
The contours with the intervals of the 5 $\sigma$ levels start from 
the 5 $\sigma$ levels, where the 1 $\sigma$ noise levels are 0.09 K km s$^{-1}$, 0.08 K km s$^{-1}$, and 0.05 K km s$^{-1}$
in ${T}^{*}_{\mathrm{A}}$ for S87E, S88B, and Gem 4, respectively.
The outlines of the blue-shifted and red-shifted outflows are plotted on the velocity dispersion maps.
The filled circle at the bottom left corner in each panel shows the effective resolution in FWHM of $27^{\prime\prime}$.
The references of outflows are Barsony (1989), Phillips $\&$ Mampaso (1991) and Snell et al. (1988)
for S87E, S88B and Gem 4.
The stellar marks indicate the position of cluster center.
The pink arrows show 1 pc scale of individual regions.}
\label{maph1}
\end{figure}

\begin{figure}
\epsscale{0.9}
\plotone{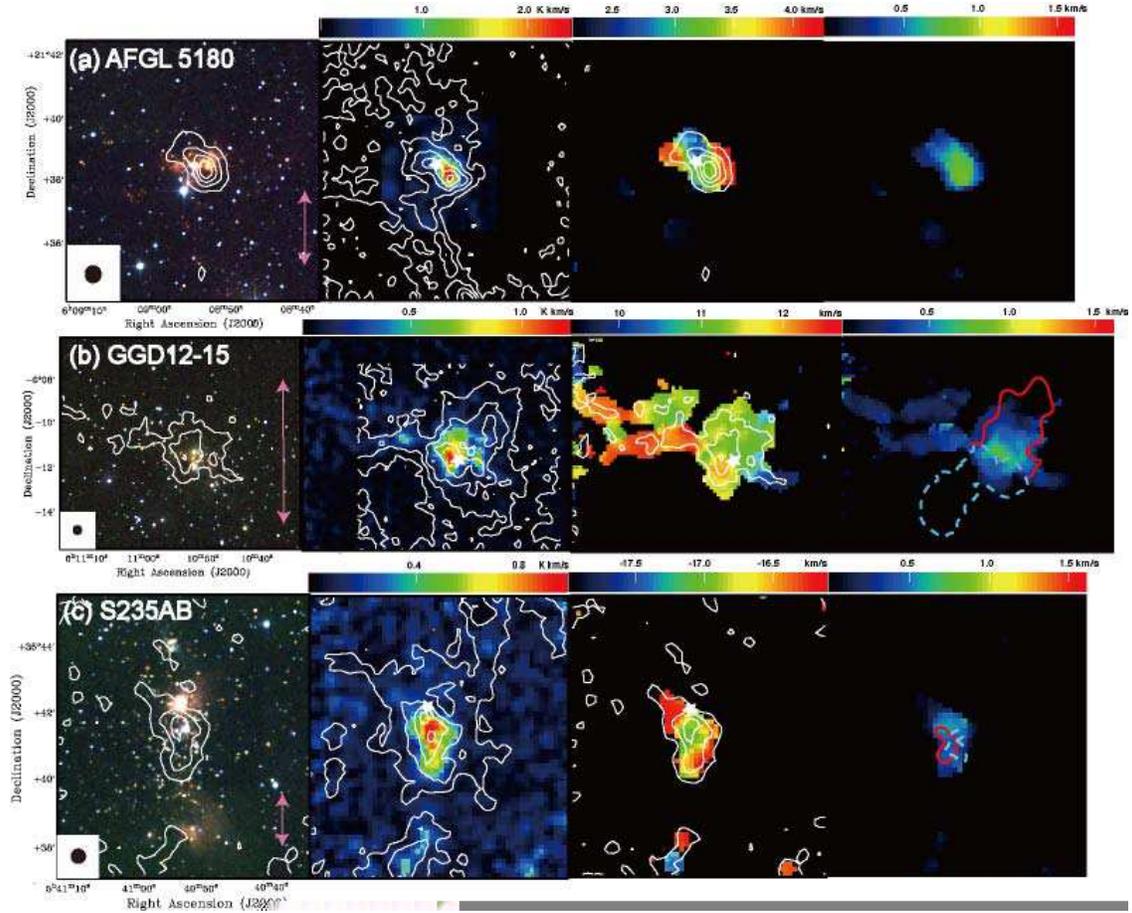}
\caption{
Same as Figure \ref{maph1} for AFGL 5180 (a), GGD12-15 (b), and S235AB (c).
The 1 $\sigma$ noise levels are 0.06 K km s$^{-1}$, 0.07 K km s$^{-1}$, and 0.04 K km s$^{-1}$ in ${T}^{*}_{\mathrm{A}}$ 
for AFGL 5180, GGD12-15, and, S235AB respectively.
The references of outflows are Little et al. (1990) and Nakano $\&$ Yoshida (1986) 
for GGD12-15 and S235AB.}
\label{maph2}
\end{figure}

\begin{figure}
\epsscale{0.9}
\plotone{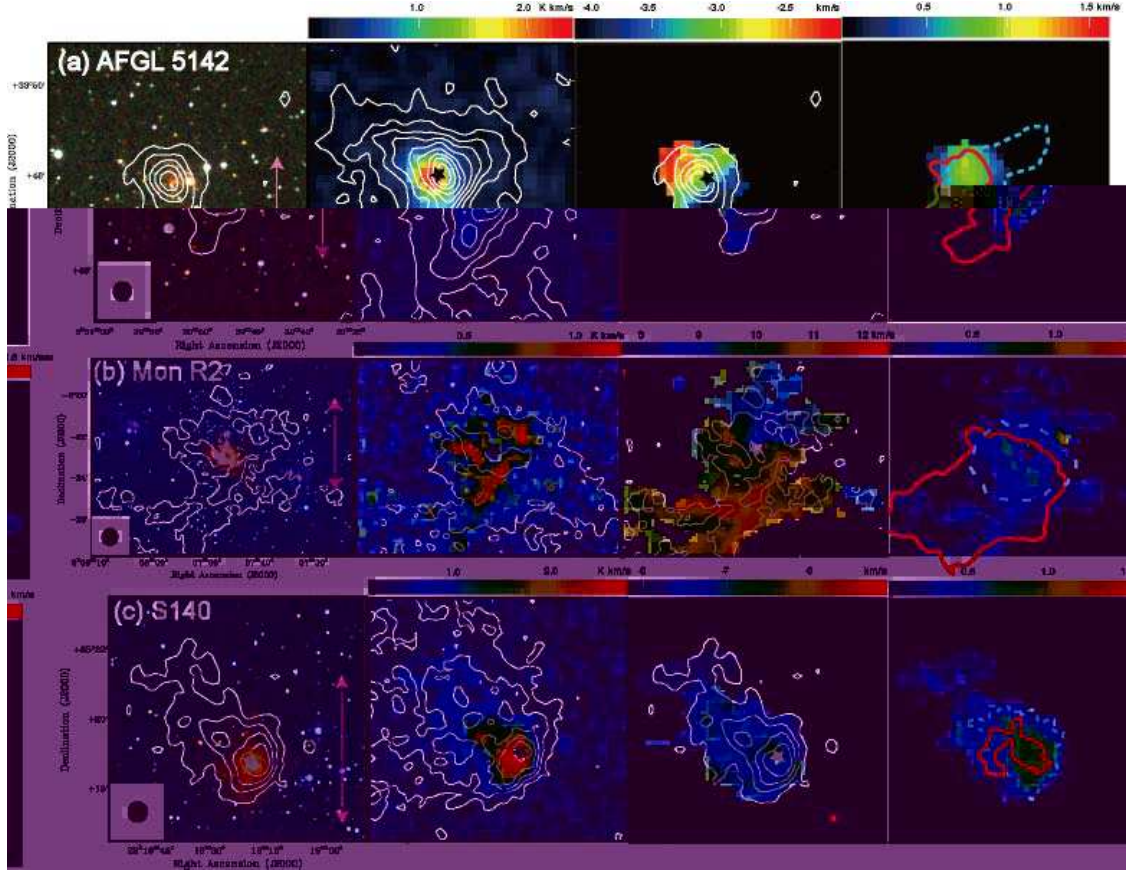}
\caption{Same as Figure \ref{maph1} for AFGL 5142 (a), Mon R2 (b) and S140 (c).
The 1 $\sigma$ noise levels are 0.07 K km s$^{-1}$, 0.07 K km s$^{-1}$, and 0.12 K km s$^{-1}$ in ${T}^{*}_{\mathrm{A}}$ 
for AFGL 5142, Mon R2, and S140, respectively.
The references of outflows are Hunter et al. (1995), Wolf et al. (1990) and Hayashi et al. (1987)
for AFGL 5142, Mon R2 and S140.
\label{maph3}}
\end{figure}

\begin{figure}
\epsscale{0.9}
\plotone{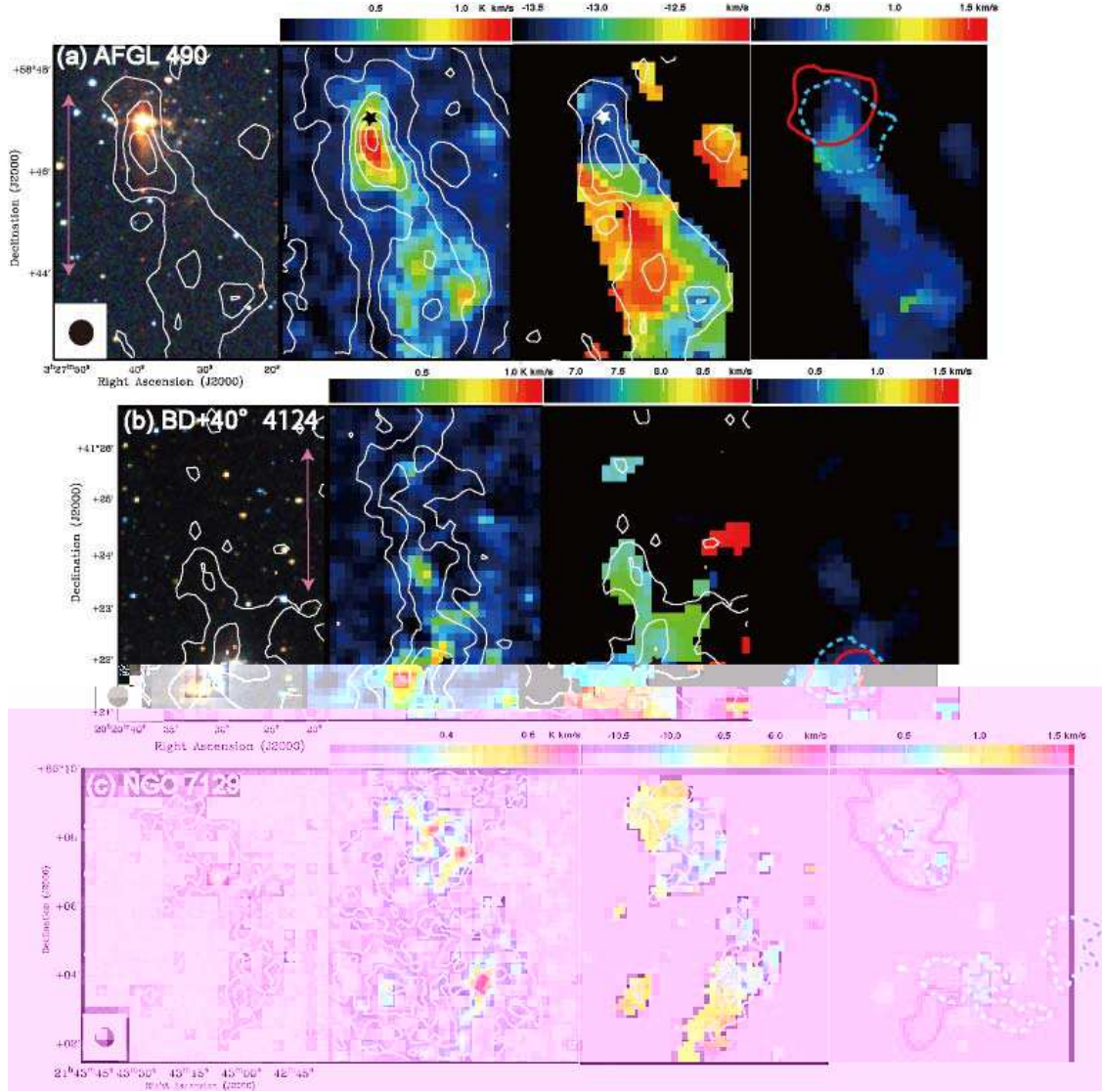}
\caption{Same as Figure \ref{maph1} for AFGL 490 (a), BD+40$^{\circ}$4124 (b) and NGC 7129 (c).
The 1 $\sigma$ noise levels are 0.06 K km s$^{-1}$, 0.08 K km s$^{-1}$, and 0.07 K km s$^{-1}$ in ${T}^{*}_{\mathrm{A}}$ 
for AFGL 490, BD+40$^{\circ}$4124, and NGC 7129, respectively.
The references of outflows are Mitchell et al. (1995), Palla et al. (1995) and Eiroa et al. (1998) 
for AFGL 490, BD+40$^{\circ}$4124 and NGC 7129.
\label{maph4}}
\end{figure}

\begin{figure}
\epsscale{0.9}
\plotone{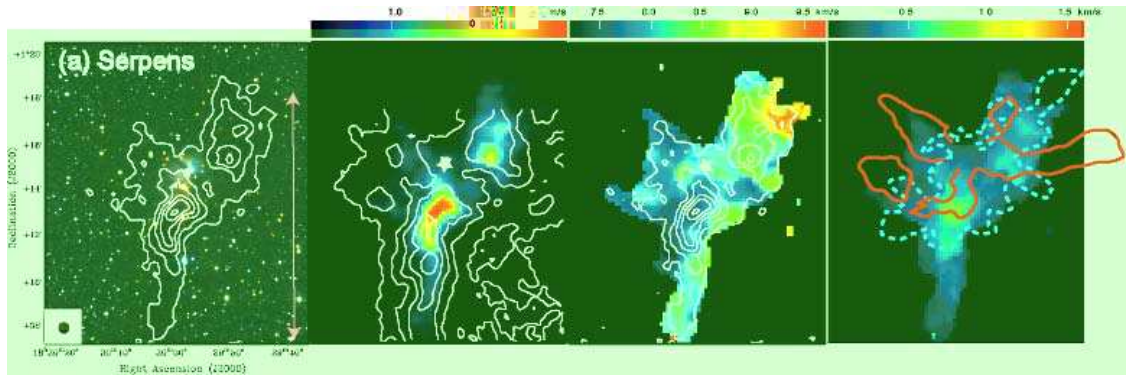}
\caption{Same as Figure \ref{maph1} for Serpens SVS2 (a). The 1 $\sigma$ noise levels are 0.06 K km s$^{-1}$ 
in ${T}^{*}_{\mathrm{A}}$.
The reference of outflows is Davis et al. (1999) for Serpens.
\label{maph5}}
\end{figure}

The left panels of Figures \ref{maph1}--\ref{maph5} show the $\HCO$ total integrated intensity maps 
superposed on the Two Micron All Sky Survey (2MASS) images ($J,H, \ $\rm{and}$ \ {K}_{\rm{s}}$ composite color images).
We defined the gas components traced by the $\HCO$($J$=1--0) line emission above the 5 $\sigma$ noise level in the individual maps as the $\HCO$ clumps.
Higuchi et al. (2009) detected the $\CO$ emission for all of the 14 objects observed in $\HCO$ in this work. 
For Gem 1, no $\HCO$ line emission was detected above the 5 $\sigma$ noise level, and hence it will not be shown in our discussion hereafter.
The detailed derivations of the physical properties are described in Section \ref{3-2}.
The $\HCO$ clumps show various morphologies, compact or extended, spherical or filamentary.
Moreover, our high-resolution maps show that most of the clumps have internal structures which consist of several emission components.
We defined the internal components whose separations are comparable to the order of Jeans length as the 
sub-clumps.\footnote{We defined the internal substructures as sub-clumps using the same method as Higuchi et al. (2009).}
The Jeans length, $\lambda_{\rm{J}}$, was calculated as 
$\lambda_{\rm J}=0.08\,{\rm pc}\left(T_{\rm K}/20\,{\rm K}\right)^{1/2}\left(n_{\rm H_2}/10^5\, {\rm cm^{-3}}\right)^{-1/2}$, 
where $T_{\rm{K}}$ is the kinetic temperature and $n_{\rm H_{2}}$ is the density of the gas components. 
We derive the kinetic temperature, $T_{\rm{K}}$ in Section \ref{3-2}.
The density of the clump, $n_{\rm H_{2}}$, was calculated as $n_{\rm H_{2}}=({3/4\pi})({\MCLU}/\mu{\rm{m_{H_{2}}}}{\RC}^{3})$, 
where $\MCLU$ is LTE mass of the clump, $\mu{\rm{m_{H_{2}}}}$ is mean mass of molecule, and $\RC$ is the radius of the clump derived in Section \ref{3-2}.
The definitions of peaks, cavities, and offset scales are given in Higuchi et al. (2009).
For each target source, we defined the cluster center as the position of the most massive star, 
the peak position as position where the $\HCO$ column density is locally enhanced by a factor of a few compared to the mean value of the clump 
($\HCO$ line emission above $\sim$ 15 $\sigma$ region in all objects). 
We can see a variation in the relative positions of the stellar clusters in the distributions of the $\HCO$ emission. 
For example, while for AFGL 5142 and S140 the cluster members are just located at the single peak position of the $\HCO$ 
emission, for Mon R2 and NGC 7129 the stellar components are not located at the emission peaks but at the emission skirts or valleys. 
The middle-right panels of Figures \ref{maph1} to \ref{maph5} show the intensity-weighted 1st moment maps of the $\HCO$ emission. 
Some of the $\HCO$ clumps are found to have relatively clear velocity gradients within the clumps, 
especially for S87E, S88B, AFGL 5180, AFGL 5142, Mon R2, and Serpens SVS2.
There appears to be a correlation between the direction of the velocity gradients and the orientation of the elongated clumps.
The right panels of Figures \ref{maph1} to \ref{maph5} show the intensity-weighted 2nd moment maps in $\HCO$. 
Most of our target cluster-forming regions were found to have molecular outflows in previous studies 
(e.g.,$\,$Wolf et al. 1990 for Mon R2) as shown in the right panels of Figure \ref{maph1} to \ref{maph5}.
The references for the outflow data are given in the captions of Figure \ref{maph1} to \ref{maph5}.

\subsection{Physical properties of the $\HCO$ clumps}\label{3-2}

In this section we describe the derivation of the physical properties of the $\HCO$ clumps.
In Table \ref{para}, we summarize the physical properties derived using the following methods.
In order to define the clump radius, $R_{\rm{clump}}$, we firstly calculated the observed radius, $R_{\rm{obs}}$.
We directly measured the projected area of the $\HCO$ clump on the sky plane $A$, 
which is the emission area defined by the 5 $\sigma$ contour (mean intensity is $\sim$ 0.8$\,$K). 
The observed radius was derived as $R_{\rm{obs}}$=$(A/\pi)^{1/2}$.
$R_{\rm{clump}}$ was estimated with correcting for broadening effect by the observing beam as

\begin{eqnarray}
\label{radius}
R_{\rm{clump}} \quad=\quad [ R_{\rm{obs}}^2 -
		(\theta_{\rm FWHM}/2)^2]^{1/2} ,
\end{eqnarray}
where $\theta_{\rm{FWHM}}$ of $27''$ is the effective beam size in FWHM.

The velocity width of the clump (one-dimensional velocity widths in FWHM), $\dvC$ was derived as follows.
We derived the $\HCO$($J$=1--0) line widths ($\dv_{\mathrm{obs}}$ in FWHM) by fitting the $\HCO$($J$=1--0) 
spectra averaged over the clumps with a Gaussian function. 
We estimated intrinsic velocity widths of the clumps, $\dvC$, by correcting for the line broadening by profile synthesis in 
auto-correlators, $\dv_{\mathrm{spec}}$, as $\dvC$ =$\left(\dv_{\mathrm{obs}}^2-\dv_{\mathrm{spec}}^2 \right)^{1/2}$. 
We found that the velocity widths, $\dvC$ range from 1.5$\,$${\rm{km \: s}^{-1}}$ to 4.0$\,$${\rm{km \: s}^{-1}}$, 
which indicates that non-thermal contribution is dominant; the thermal velocity widths 
(one-dimensional velocity widths in FWHM) of $\HCO$ molecules are 0.08--0.14$\,$${\rm{km \: s}^{-1}}$ with $T_{\rm{K}}$ = 10--30$\,$K.
The H$_{2}$ column density of the clumps, $N(\rm{H}_{2})$ was calculated as,
\begin{eqnarray}
\label{column}
{N(\rm{H}_{2})} = 1.0 \times 10^{21} \left(\frac{X_{\HCO}}{5.6\times 10^{-11}}\right)^{-1}
	\left(\frac{T_{\rm{ex}}}{\rm{K}}\right)
	\exp \left({\frac{4.16}{T_{\rm{ex}}/\rm{K}}}\right)
	\nonumber \\
	\quad \quad \quad
	\times	
	\left( \frac{\eta}{0.51} \right)^{-1}
	\left( \frac{\tau}{1-\exp({-\tau})} \right)
 	\left( \frac{\int {{T}^{*}_{\rm{A}}} dv}{\rm{K} \: \rm{km \: s}^{-1}} \right)
	\quad [\rm{cm}^{-2}] ,
\end{eqnarray}
where $X_{\HCO}$ is the fractional abundance of $\HCO$ relative to $\rm{H}_2$, 
$T_{\mathrm{ex}}$ is the excitation temperature of the transition, $\tau$ is the optical depth,
and $\int {T}^{*}_{\mathrm{A}} dv$ is the total integrated intensity of the $\HCO$ line emission.
For $X_{\HCO}$, we used 5.6 $\times$ $10^{-11}$ \citep{aoy01}.
Here, we applied the optically thin condition ($\tau$ $\lesssim$ 1) in $\HCO$ line emission because
we found that the $\HCO$ line emission peak is weaker than the peak intensity of $\CO$ line emission 
which is considered to be optically thin \citep{hig09}.
In order to estimate the excitation temperature of $\HCO$, $T_{\rm{ex}}$, we used the NH$_{3}$ data from the NRO data archive,
which have already been presented in Higuchi et al. (2009).
Under the local thermodynamic equilibrium (LTE) condition, we consider that the excitation temperature 
is comparable to the kinetic temperature ($T_{\rm{ex}}$ $\sim$ $T_{\rm{K}}$).
We applied the relation between the kinetic temperature and the rotation temperature, $T_{\rm{K}}$ $\sim$ $T_{\rm{rot}}$ 
as proposed by Danby et al. (1988).
The rotation temperatures, ${T_{\rm{rot}}}$(2,2;1,1) were derived from the NH$_{3}$ data using the same method 
presented by Ho $\&$ Townes (1983).

The total mass under the LTE condition, $\MCLU$ is calculated as,
\begin{eqnarray}
\label{mass}
\MCLU = 15 \left(\frac{X_{\HCO}}{5.6\times 10^{-11}}\right)^{-1}
	\left( \frac{D}{2100 \: \mathrm{pc}} \right)^{2} 
	\left( \frac{\RC}{50 ''} \right)^{2} 
	\nonumber \\
	\quad \quad \quad
	\times
	\left(\frac{T_{\rm{ex}}}{\rm{K}}\right)
	\exp \left({\frac{4.16}{T_{\rm{ex}}/\rm{K}}}\right)
	\left( \frac{\eta}{0.51} \right)^{-1}
 	\left( \frac{\int {{T}^{*}_{\rm{A}}} dv}{\rm{K} \: \rm{km \: s}^{-1}} \right)
	\quad [\MO] ,
\end{eqnarray}
where $D$ is the distance to the object and $\RC$ is the radius of the clump. 
Higuchi et al. (2009) calculated the virial mass of the clumps under 
the condition that the internal energy of the clumps is contributed by the thermal and turbulent motions of cloud constituent. 
However, this condition is likely invalid for the $\HCO$ clumps because some of those have distinct velocity gradients. 
Instead, we discuss the energy balance of the $\HCO$ clumps by taking into account the contribution of rotation (see Section \ref{4-2}).

\subsection{Comparison between the physical properties of the $\CO$ clumps and the $\HCO$ clumps}\label{3-3}

In this section, we focus on the spatial relations, and the comparison of the physical parameters between the $\CO$ and $\HCO$ clumps.
The middle-left panels of Figure \ref{maph1} to \ref{maph5} show the total integrated intensity maps of 
the $\CO$($J$=1--0) emission (contours; Higuchi et al. 2009) superposed on the $\HCO$ emission.
By comparing the spatial distribution between $\CO$ and $\HCO$ line emission in each object, 
we found that $\HCO$ emission is concentrated where the $\CO$ line emission is strong and
the $\HCO$ clumps are found to have compact and simpler shapes than the $\CO$ clumps, which suggests that the $\HCO$
line emission traces dense region within the $\CO$ clumps. 
For example, there are very compact $\HCO$ clumps within the extended $\CO$ clumps in S88B.
Mon R2 and NGC 7129 which have complex structures in $\CO$ emission have compact multiple emission components in $\HCO$. 
For most of the observed objects, the positions of the emission peak of $\HCO$ agree with those of $\CO$. 
Some of them, S87E and Serpens SVS2 have emission peaks of $\HCO$ which are in disagreement with
those of $\CO$ by an order of Jeans length.

By comparing the LTE mass between the $\CO$ and the $\HCO$ clumps, 
we found that the ratio of $\MCLU(\HCO)$ to $\MCLU(\CO)$ is below unity in 90$\,\%$ of our targets.
Only for AFGL 5180, the ratio of $\MCLU(\HCO)$ to $\MCLU(\CO)$ becomes unity. 
Comparison of the velocity widths between $\CO$ clumps and $\HCO$ clumps suggests that
the ratio of $\dv(\HCO)$ to $\dv(\CO)$ is below unity in half of our targets.
The rest (50$\,\%$) of our targets with relatively massive stars tend to have larger ratios than unity.
They have the large $\dv(\HCO)$ ($\gtrsim$ 2 km s$^{-1}$) values for S87E, S88B, Gem 4, AFGL5180, AFGL5142, and S140. 
Two thirds (except for Gem 4, S140) of our target clumps with large $\dv(\HCO)$ have clear velocity gradients in the 1st moment maps.
The detailed parameters are listed in Table \ref{class}.

\section{DISCUSSION}

We observed the $\HCO$ emission toward 14 embedded clusters in order to address the formation and evolution of clusters. 
In this section, we discuss the evolutionary status and the formation mechanism of clusters from the observed 
physical properties in terms of the following aspects: 
(1) evolution of dense gas and star formation efficiency, (2) kinematic structure of dense clumps, and 
(3) clump-clump collision as a possible triggering of cluster formation.

\subsection{Evolution of dense gas in cluster-forming regions}\label{4-1}

\subsubsection{Morphological classification of $\HCO$ clumps}\label{4-1-1}

In this section, we focus on the morphological features of $\HCO$ clumps and 
the relation of the spatial distribution between $\HCO$ clumps and stellar clusters. 
Our observations revealed that the $\HCO$ clumps show various morphologies, compact (e.g.,$\,$AFGL 5142) or extended structure 
(e.g.,$\,$Mon R2), cometary (e.g.,$\,$S140) or filamentary (e.g.,$\,$Gem 4, Serpens SVS2).
Most of them have single or double components within the clump.
We categorize the $\HCO$ peak positions into two cases, ``Peak" and ``Off-peak", 
using the separation between the $\HCO$ peak positions and the cluster centers
; if the separation is smaller than the Jean length estimated for each target, we call them the ``Peak" or instead of the ``Off-peak". 
The number of identified sub-clumps is dependent on the spatial resolution of the observations and the distance of the target objects. 
Note that for a distant object, two or more sub-clumps might be identified as one component of sub-clump. 
Nevertheless, for S87E at the distance of 2.1$\,$kpc, the separation between the sub-clumps corresponds to $\sim$ 0.3$\,$pc 
which is twice of Jeans length. Hence, the above discussion still holds with our definition and identification of the sub-clumps.
Additionally, the $\HCO$ maps show a variation of brightness contrast.
Under the optically thin condition for the $\HCO$ line, we can assume that the brightness contrast is equivalent to the column density contrast, 
which reflects the physical state of the dense regions within the clumps.
There is a variation in the relation of the spatial distribution between $\HCO$ peaks and cluster centers. 
Extreme cases are found in two sample groups; a stellar cluster located just at a single peak in the distribution of the $\HCO$ line emission 
(e.g.,$\,$AFGL 5142, S140), or located at a cavity of the emission line distribution within the clump which has multiple components around the cavity 
(e.g.,$\,$Mon R2). As an intermediate case between them, a cluster is associated with one of the $\HCO$ peaks 
(e.g.,$\,$AFGL 490) or not associated with $\HCO$ emission peak, but located within the clumps (e.g.,$\,$S87E). 
There are two distinct clusters in Gem 4, and one of them is located at the emission peak of $\HCO$.

Higuchi et al. (2009) proposed a morphological classification using the $\CO$ emission line for the same targets.
We applied the same classification scheme to the $\HCO$ clumps and sort the clumps into three types to simplify discussion 
and to clarify a qualitative trend.
We define $\it{Type \ A}$ to be clumps in which the clusters are just associated with a single peak of $\HCO$ emission distribution. 
These clumps not only have a single peak but a higher brightness contrast in the distribution of the $\HCO$ emission. 
We define $\it{Type \ C}$ to be clumps in which the clusters are located apart from the main $\HCO$ emission, or at a cavity-like $\HCO$ emission hole. 
The clumps of this type have lower brightness contrast than $\it{Type \ A}$.
The rest are $\it{Type \ B}$ clumps in which the cluster is associated with one of the peaks of the $\HCO$ emission distribution, or not 
associated with $\HCO$ emission peak but located within the clumps. 
The clumps of this type have relatively higher brightness contrasts than $\it{Type \ C}$.
We classified AFGL 5142 and S140 into $\it{Type \ A}$.
In contrast, we classified S88B, S235AB, Mon R2, NGC 7129, and Serpens SVS2 into $\it{Type \ C}$.
In addition, we classified S87E, Gem 4, AFGL 5180, GGD12-15, BD+40$^{\circ}$4124, and AFGL 490 into $\it{Type \ B}$.
The detailed parameters of individual sources are listed in Table \ref{class}.

\begin{figure}
\epsscale{0.8}
\plotone{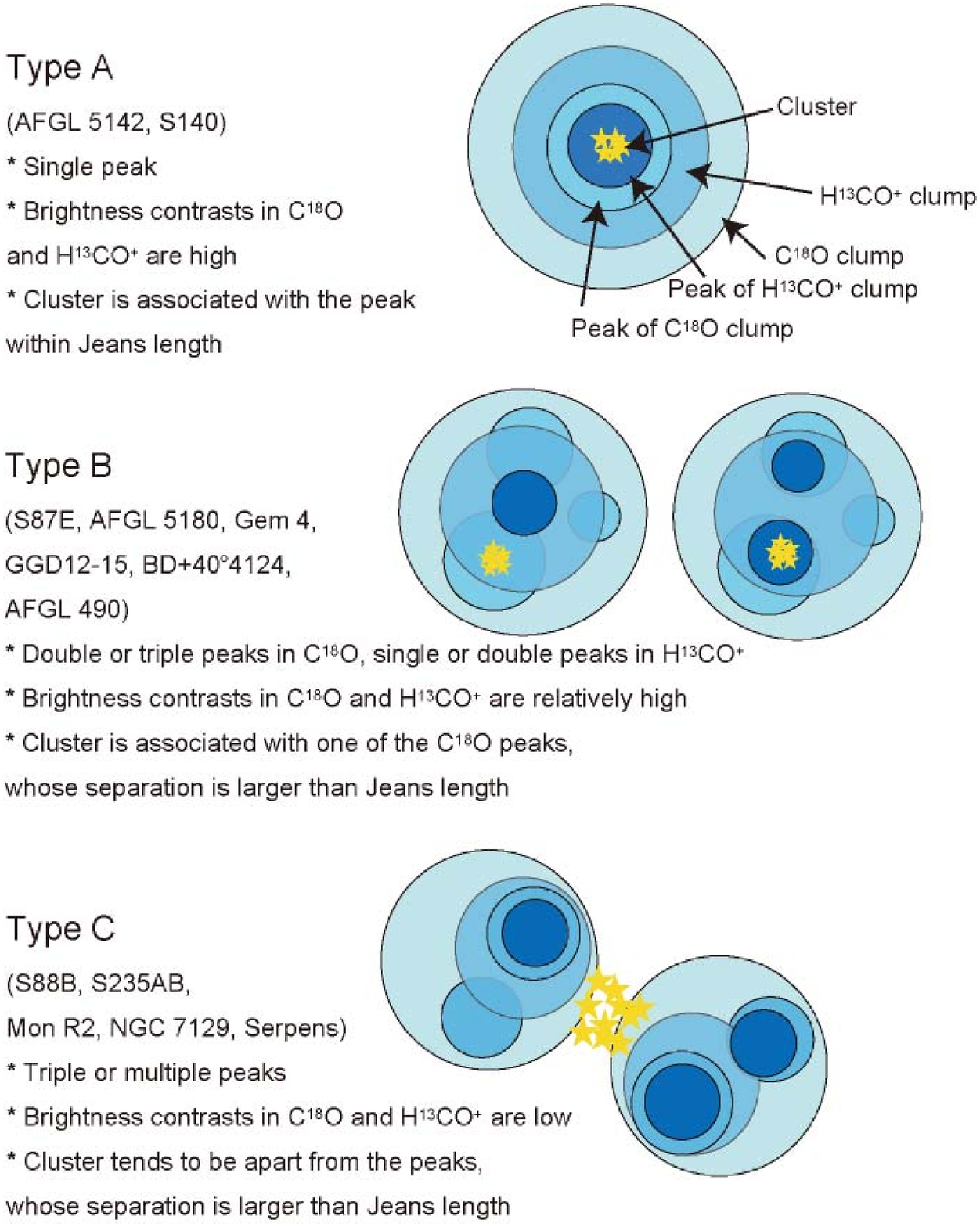}
\caption{Schematic picture of our morphological classification of cluster-forming clumps.
$\it{Type \ A}$ clumps in which the clusters are just associated with a single peak of $\CO$ and $\HCO$ emission distribution. 
$\it{Type \ B}$ clumps in which cluster is associated with one of the peaks of $\CO$ emission distribution, and 
associated with one of the $\HCO$ emission peaks. 
Some of them are not associated with $\HCO$ emission peak, but located within the $\HCO$ clump. 
$\it{Type \ C}$ clumps in which the clusters are located at a cavity-like $\CO$ and $\HCO$ emission hole. 
Individual sources are classified into the same stages in $\CO$ \citep{hig09}.
The classifications are interpreted as the evolution of cluster-forming clumps from $\it{Type \ A}$ to $\it{C}$.
\label{model}}
\end{figure}

We found that the above stages of the $\HCO$ clumps are consistent with those of the $\CO$ clumps except for AFGL 5180, and Gem 1.
This result shows that the dense regions of the clumps are extracted by the $\HCO$ line emission more than those traced by the $\CO$ line emission. 
As in Higuchi et al. (2009), we confirmed that the evolutionary sequence from $\it{Type \ A}$ to $\it{C}$ progresses with dispersing the $\HCO$ clumps.
For AFGL 5180, the classes are different depending on which line is used; $\it{Type \ C}$ ($\CO$) and $\it{Type \ B}$ ($\HCO$). 
The reason for classifying this object as $\it{Type \ C}$ is mainly because of the multiple peaks of the $\CO$ clump and the cluster located between 
them. On the other hand, the $\HCO$ clump has a single peak, thus, it is classified into $\it{Type \ B}$, which is robust for 
a dense region of cluster-forming clump because of the consistency of distribution in the $\HCO$ emission with that in 
dust continuum emission \citep{kle05}. We suggested that the classes which represent the evolutionary stages 
of cluster-forming clumps are more suitable to be determined from the $\HCO$ data than from the $\CO$ data. 
The clear trend in the relation of classes and SFEs from the $\HCO$ data shown in Section \ref{4-1-2} also agrees with this suggestion.
For Gem 1, we did not detect $\HCO$ emission although we detected and identified the $\CO$ clump, which was classified as $\it{Type \ B}$. 
From the map of dust continuum emission \citep{kle05}, it is understood that the emission component of dense region is too weak and compact 
to be detected significantly even with our observations. We estimated an upper limit to the dense gas fraction defined by 
$\MCLU(\HCO)$/$\MCLU(\CO)$ with the 5 $\sigma$ noise level of the $\HCO$ map, 
and found it to be 0.01, an order of magnitude smaller value than those of other objects. 
Thus, we consider that Gem 1 is an exceptional case, in which the dense gas might be regulated to form or dispersed effectively. 
A schematic picture of these three types using the spatial distributions of $\CO$ and $\HCO$ is shown in Figure \ref{model}.

\subsubsection{Star formation efficiency variation along the clump-cluster evolution}\label{4-1-2}

We proposed the evolutionary scenario of dense clumps associated with embedded clusters in Higuchi et al. (2009).
Since a large amount of gas is dispersed by the cluster members during cluster evolution, stellar mass becomes more dominant in the clump-cluster system.
We found a tendency along the type sequence $\it{A}$-$\it{C}$ for the clump structures to become complex having some sub-clumps and local peaks. 
The typical separation among the sub-clumps is the order of the Jeans length, which supports the following interpretation. 
In molecular clumps, the column density structure is more or less uniform initially. 
Eventually, fragmentation occurs with some mechanisms in the clumps and initiates star cluster formation \citep[e.g.,][]{bon03}. 
After fragmentation, the local condensations grow to become distinct peaks where stellar clusters will eventually be born. During the cluster formation, the cores within the clumps form stars in a wide mass range, and dense gas is dispersed by their stellar activities. 
Once clusters are formed, young stars, particularly the most massive stars, 
disperse surrounding gas with evolution making cavities around them \citep[e.g.,][]{fue98,fue02}. 
The structures of the evolved clumps are expected to be less condensed than those of the younger clumps. 
For more evolved clumps, continuous dispersal of the dense gas around the stellar cluster makes a disagreement 
between the stellar positions and the peaks of gas condensations in which clusters are located at the cavity-like structure of the clumps.

\begin{figure}
\epsscale{0.7}
\plotone{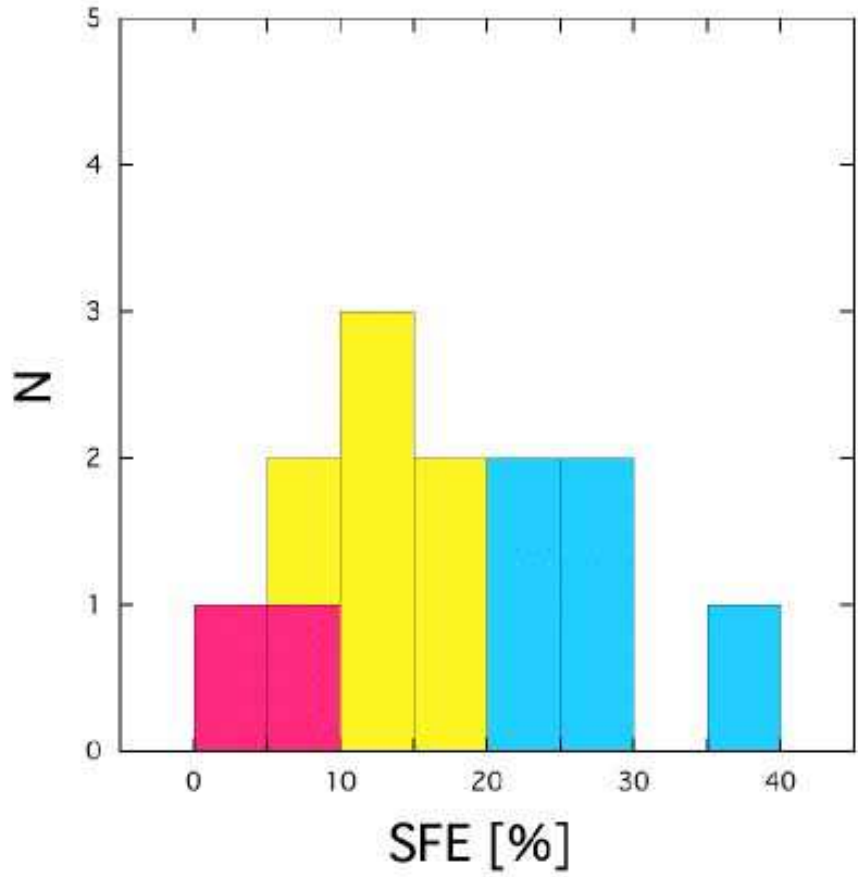}
\caption{Histogram of the SFEs of $\HCO$ clumps. 
The pink color shows $\it{Type \ A}$ object, the yellow color shows $\it{Type \ B}$ object, 
and the blue color shows $\it{Type \ C}$ object.}
\label{sfe}
\end{figure}

From our proposed scenario, the star formation efficiency
(SFE = ${M_{\rm{cluster}}}$/$(M_{\rm{cluster}}+M_{\rm{clump}})$ where $M_{\rm{cluster}}$ is the stellar masses of clusters, which are listed in Lada $\&$ Lada (2003) and $M_{\rm{clump}}$ is the masses of clumps) is expected to increase from $\it{Type \ A}$ to $\it{Type \ C}$. 
Only for S140 and BD+40$^{\circ}$4124, the total masses of the stellar members are estimated using the $N_{\rm{star}}$-$M_{\rm{cluster}}$ relation that was derived using the data in Lada $\&$ Lada (2003).
We calculated the SFEs for our targets and listed the results in Table \ref{para}.
We found that the SFEs for the $\HCO$ clumps range from 3$\,$$\%$ to 35$\,$$\%$ with a mean of 20$\,$$\%$,
which is larger than the values previously derived for low-mass star forming regions 
(e.g.,$\,$Onishi et al. 1998;$\,$Tachihara et al. 2002 for the results of $\CO$ line emission).
Figure \ref{sfe} shows a histogram of the SFEs of each type and indicates that $\it{Type \ A}$ objects have smaller 
SFEs ($\la$ 8$\%$) than $\it{Type \ B}$ and $\it{Type \ C}$ objects ($\ga$ 10$\%$).
We found on a similar trend in $\CO$ data; the larger SFEs are derived along the classes from $\it{Type \ A}$ to $C$.
The velocity gradients for DVSOs appear clearer in $\HCO$ than in $\CO$.
In previous studies, the correlation between changing of morphologies and SFE along evolution was suggested \citep[e.g.,][]{lad03,tho06,gut09}.
Higuchi et al. (2009) demonstrated a modest correlation between morphological variation and SFEs. 
In this paper, we confirmed that the correlation is clearer with $\HCO$ than $\CO$ (see Figure 8 (b) in Higuchi et al. (2009)).
This result indicates that $\HCO$ traces the dense regions in the $\CO$ clumps presented in Higuchi et al. (2009) 
and physical conditions of dense regions are sensitive and affected by the clump-cluster evolution.

\subsection{Kinematic structures of the clumps}\label{4-2}

We found that half of our objects (S87E, S88B, AFGL 5180, AFGL 5142, Mon R2, and Serpens SVS2) have clear velocity gradients 
in the 1st moment maps as shown in the middle-right panels of Figures \ref{maph1} to \ref{maph5}.
The others have random velocity structures in the 1st moment maps.
For the $\CO$ clumps of the same targets presented in Higuchi et al. (2009), none of them have clear velocity gradient in the 1st moment maps.
The $\CO$ line emission is considered to trace not only the dense region but also surrounding gas of the dense region, 
and to barely clarify the velocity field in the dense region. 
We suggest that the velocity gradient is one of the most important features of the dense region in the cluster-forming clumps 
for the understanding of the evolution of clumps that form stellar clusters.

\begin{figure}
\epsscale{1.0}
\plotone{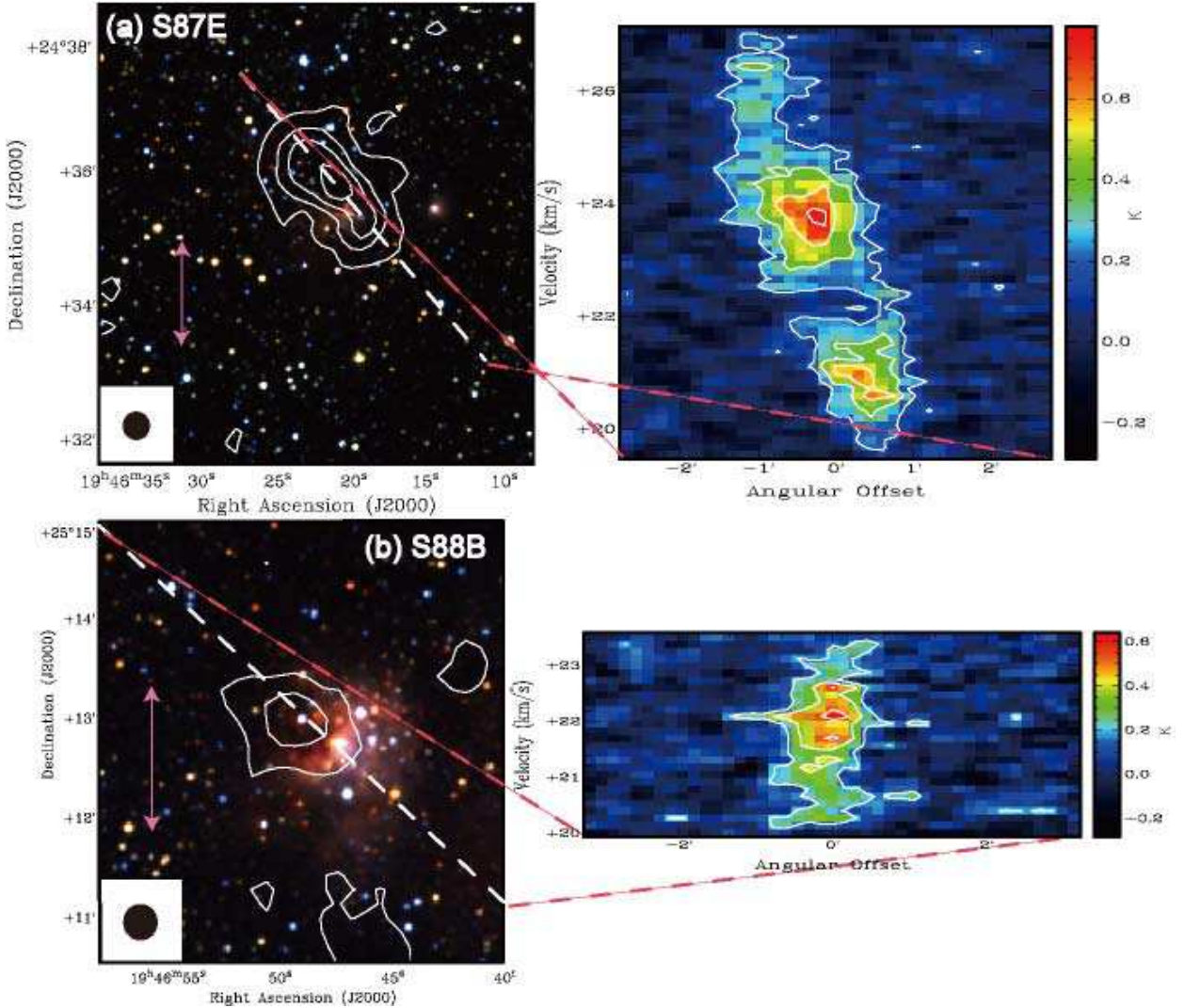}
\caption{Total integrated intensity maps of the $\HCO$($J$=1--0) emission 
(contours) superposed on the $JHK_\mathrm{s}$ composite color images in log scale from 2MASS for S87E (a) and S88B (b), which are the same as the 
left panels of Figure \ref{maph1} (left).
The right panels are the position-velocity (P-V) diagrams of $\HCO$($J$=1--0) emission.
The contours in the P-V diagrams with the intervals of the 3 $\sigma$ levels start from 
the 3 $\sigma$ levels, where the 1 $\sigma$ noise levels are 0.06 K, and 0.06 K in ${T}^{*}_{\mathrm{A}}$ for S87E and S88B.
The pink arrows show 1 pc scale of individual regions. 
\label{pvmap1}}
\end{figure}

\begin{figure}
\epsscale{1.0}
\plotone{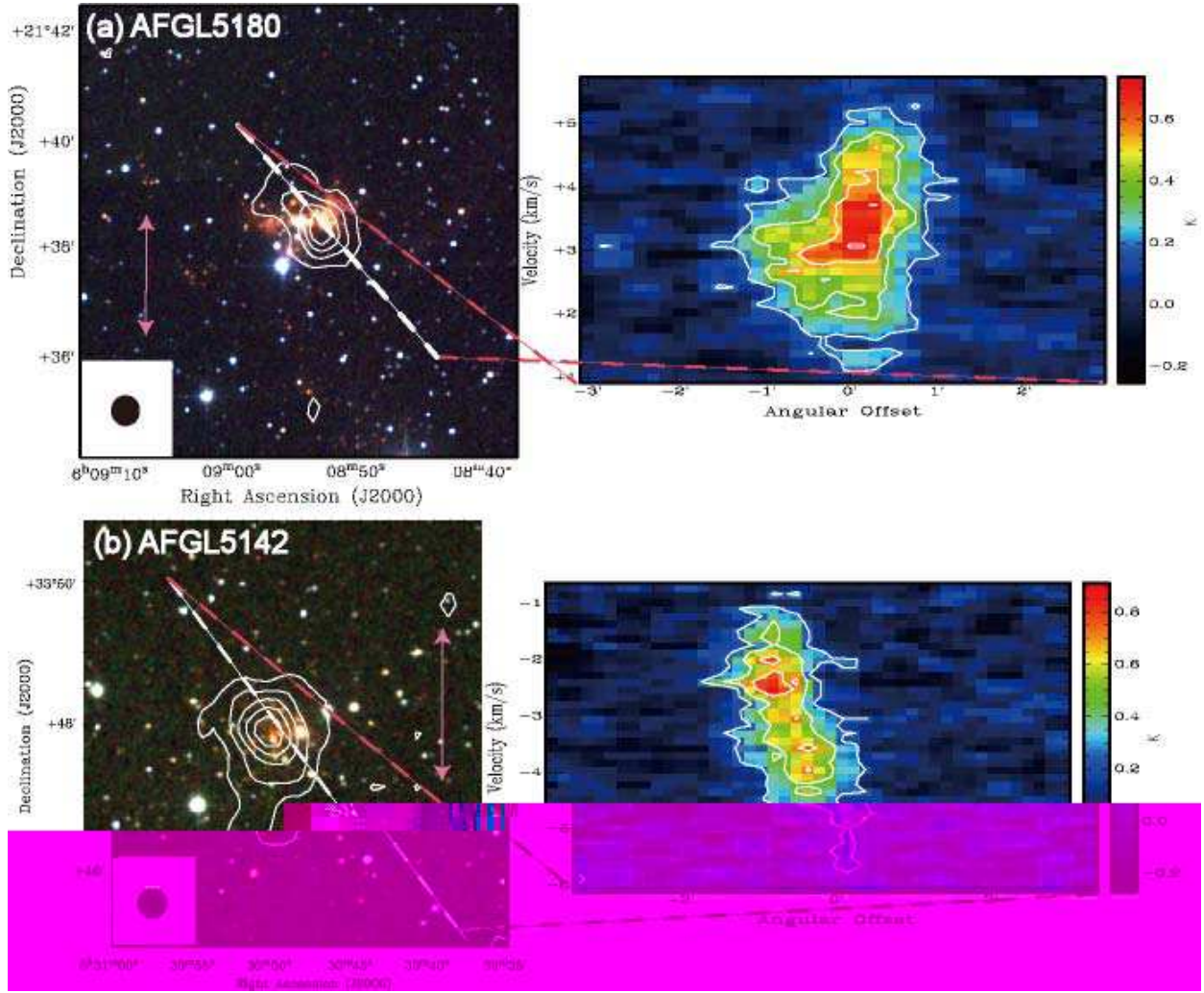}
\caption{Same as Figure \ref{pvmap1} for AFGL 5180 (a) and AFGL 5142 (b).
The contours in the P-V diagrams with the intervals of the 3 $\sigma$ levels start from 
the 3 $\sigma$ levels, where the 1 $\sigma$ noise levels are 0.06 K, and 0.08 K in ${T}^{*}_{\mathrm{A}}$ for AFGL 5180 and AFGL 5142.
\label{pvmap2}}
\end{figure}

\begin{figure}
\epsscale{1.0}
\plotone{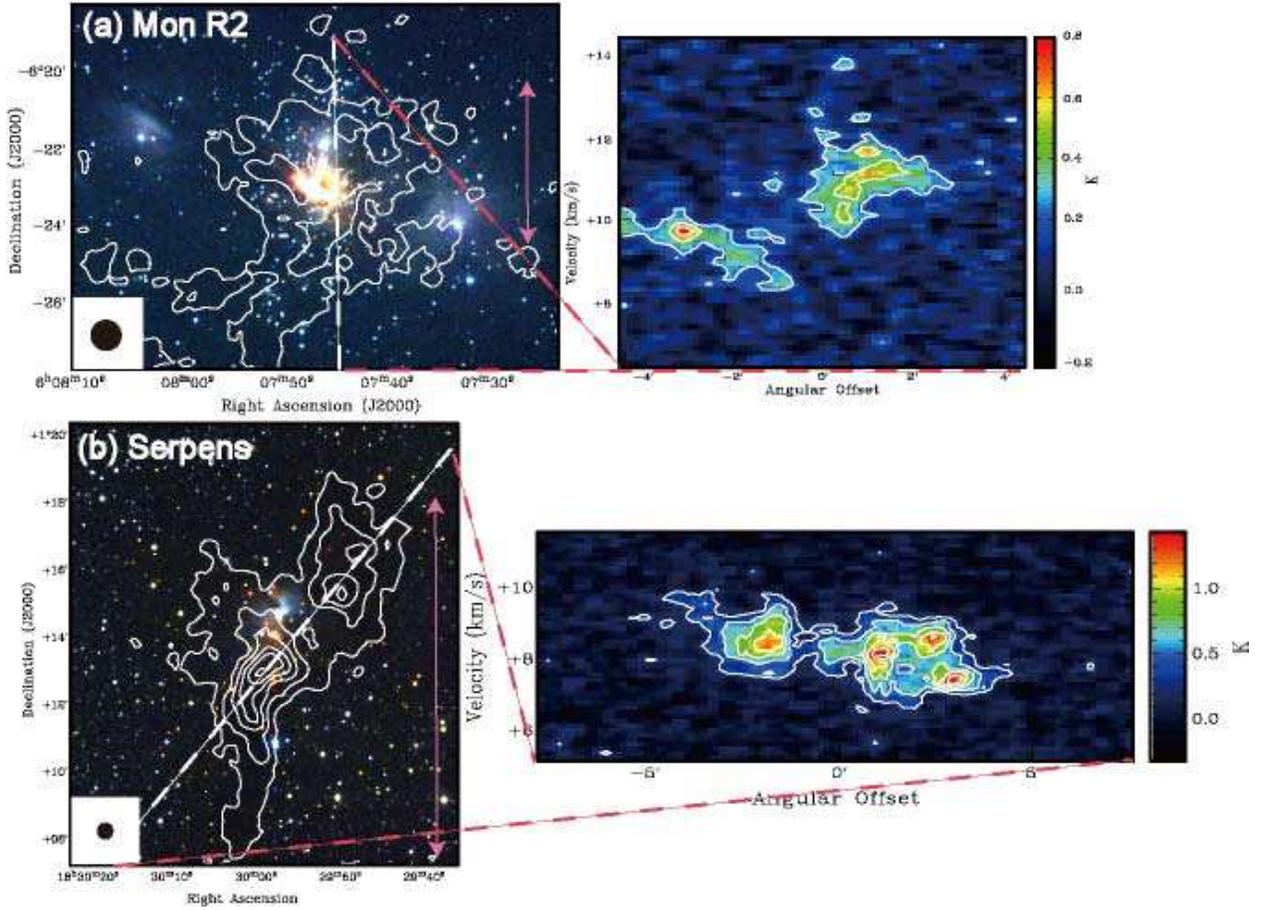}
\caption{Same as Figure \ref{pvmap1} for Mon R2 (a) and Serpens SVS2 (b).
The contours in the P-V diagrams with the intervals of the 3 $\sigma$ levels start from 
the 3 $\sigma$ levels, where the 1 $\sigma$ noise levels are 0.07 K, and 0.08 K in ${T}^{*}_{\mathrm{A}}$ for Mon R2 and Serpens SVS2.
\label{pvmap3}}
\end{figure}

\begin{figure}
\epsscale{0.8}
\plotone{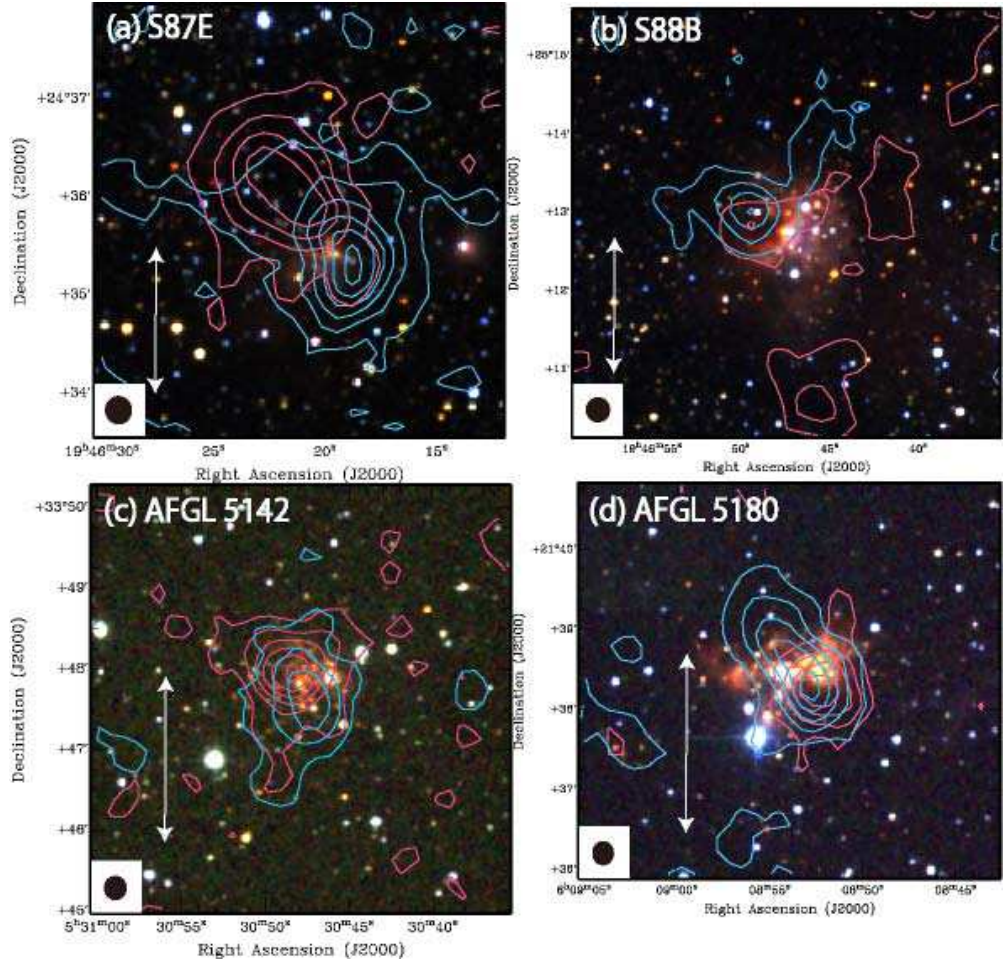}
\caption{
The maps of blue-shifted and red-shifted velocity range of the $\HCO$($J$=1--0) emission 
(contours) superposed on the $JHK_\mathrm{s}$ composite color images in log scale from 2MASS for S87E (a), S88B (b), AFGL 5142 (c), and AFGL 5180 (d).
The contours with the intervals of the 3 $\sigma$ levels start from the 3 $\sigma$ levels.
The 1 $\sigma$ noise levels for the blue-shifted range are 0.09$\,$K$\,$km$\,$s$^{-1}$, 0.06$\,$K$\,$km$\,$s$^{-1}$, 0.05$\,$K$\,$km$\,$s$^{-1}$, and 0.04$\,$K$\,$km$\,$s$^{-1}$, and those for red-shifted range are 0.07$\,$K$\,$km$\,$s$^{-1}$, 0.06$\,$K$\,$km$\,$s$^{-1}$, 0.08$\,$K$\,$km$\,$s$^{-1}$, and 0.07$\,$K$\,$km$\,$s$^{-1}$in ${T}^{*}_{\mathrm{A}}$ for S87E, S88B, AFGL 5142, and AFGL 5180, respectively. 
The white arrows show 1 pc scale of individual regions.
The velocity ranges of blue-shifted components are 20$\,$km$\,$s$^{-1}$ to 22$\,$km$\,$s$^{-1}$, 20$\,$km$\,$s$^{-1}$ to 22$\,$km$\,$s$^{-1}$, $-$5$\,$km$\,$s$^{-1}$ to $-$3$\,$km$\,$s$^{-1}$, 1$\,$km$\,$s$^{-1}$ to 4.2$\,$km$\,$s$^{-1}$ for S87E, S88B, AFGL 5142, and AFGL 5180.
The velocity ranges of red-shifted components are 22$\,$km$\,$s$^{-1}$ to 27$\,$km$\,$s$^{-1}$, 22$\,$km$\,$s$^{-1}$ to 23$\,$km$\,$s$^{-1}$, $-$1$\,$km$\,$s$^{-1}$ to $-$3$\,$km$\,$s$^{-1}$, 4.2$\,$km$\,$s$^{-1}$ to 5.4$\,$km$\,$s$^{-1}$ for S87E, S88B, AFGL 5142, and AFGL 5180.}
\label{maprb}
\end{figure}

In order to investigate the velocity structures of the $\HCO$ clumps in detail, we made position-velocity (P-V) diagrams
for six objects which have relatively clear velocity gradients in the 1st moment maps.
The axis of the slice is parallel to the direction of velocity gradients (see Figure \ref{pvmap1} to \ref{pvmap3}).
There are the bulk linear gradients which include the multiple components in P-V diagram for the four objects out of six. 
The velocity gradient in the $\HCO$ clumps were derived for all 13 objects with a linear fitting to the emission in the P-V diagrams, 
which are clipped at the 3 $\sigma$ noise levels in the channel maps.
The derived values of the velocity gradients, $\sigma_{\rm{grad}}$, are listed in Table \ref{beta}. 
The gradients range from 0.5 km$\,$s$^{-1}$$\,$pc$^{-1}$ to 4.3 km$\,$s$^{-1}$$\,$pc$^{-1}$.
The gradients for the four objects are as large as 2.2 km$\,$s$^{-1}$$\,$pc$^{-1}$ to 4.3 km$\,$s$^{-1}$$\,$pc$^{-1}$.
In addition, Figure \ref{maprb} shows the spatial distribution of the blue-shifted and red-shifted components with the clusters, 
and all of the clusters are found to be located in the overlapped areas between blue-shifted and red-shifted components; 
a typical example is S87E, showing two spatially separated clumps at both sides of the cluster center.
The velocity ranges of the blue-shifted and red-shifted components are described in the caption of Figure \ref{maprb}.
The spatial distributions of the blue-shifted and red-shifted components imply the molecular outflows. 
The molecular outflows in our targets have already been detected in previous studies. We found the following features. 
(1) There are no correlations between the orientation of outflows and those of the velocity gradients. 
(2) These outflows have been detected in the optically thick tracers (e.g., CO lines), in contrast we expected that the line emission is optically thin. 
(3) In the spectrum data, there are no wing components. 
From the above observational evidence, we conclude that the velocity structures are not made by outflow activities. 
There is a feature that the four objects (S87E, S88B, AFGL 5180, and AFGL 5142) with the above velocity structures contain relatively massive stars 
($\ga$ 10$\,$$\MO$) within the clusters. Here, we named ``DVSOs" (Distinct Velocity Structure Objects) 
after the objects with distinct velocity structure for the above objects.
These interesting kinematic structures are discussed in Section \ref{4-3} in terms of the clump-clump collision model.

We examined the energy balance of the clumps using the equilibrium virial theorem (e.g., Goodman et al. 1993).
The gravitational potential energy is estimated from $\Omega=-{\alpha_{\rm{vir}}}G\MCLU^{2}/\RC$, where $\alpha_{\rm{vir}}$ 
is the coefficient $\alpha_{\rm{vir}}=(3-p)/(5-2p)$ with the index of density profile, $p$ for the correction for derivation from constant density \citep{wil94}. 
In the assumption of Goodman et al. (1993), they adopted $p=2$ using the isothermal collapsing core models of the radial profile of the core. 
The center position of radial profile is defined as the position of star within the core. 
In contrast, the clump has inner components (e.g., sub-clumps) so that the density profile is not as simple as that of the core. 
It is difficult for us to define the center position of radial profile toward the cluster-forming clump. In this paper, we assume the uniform density sphere to describe the inner density structure of the clumps.
Assuming that the linear velocity gradient in each clump represents rotational motion, we estimated the rotational energies 
from $E_{\rm{rot}} = {\frac{1}{2}}{\alpha_{\rm{rot}}}\MCLU{\omega^{2}}\RC^{2}$, where $\alpha_{\rm{rot}}=2/(5+2p)$, 
$\omega$ is the angular velocity. We estimated the turbulent energies as, $E_{\rm{turb}}={\frac{3}{2}}{\MCLU}(\Delta{V}^{2}_{\rm{turb}}/{8\ln{2}})$, and the turbulent velocity widths $\Delta{V}_{\rm{turb}}$ were derived by subtracting $\Delta{V}_{\rm{therm}}$ and the velocity gradients from the observed velocity widths. By dividing the virial equation by $|\Omega|$, we calculated $\beta_{\rm{vir}} = 2(\beta_{\rm{therm}} + \beta_{\rm{turb}} + \beta_{\rm{rot}})-1$, $\beta_{\rm{vir}}$ is the total energy of system divided by $|\Omega|$, and $\beta_{\rm{therm}}$, $\beta_{\rm{turb}}$, and $\beta_{\rm{rot}}$ are the virial ratios of the thermal, turbulent, and rotational energies, respectively. 
Especially, it is noteworthy to compare the virial ratios of the rotational energies, which are listed in Table \ref{beta}.

From the observational data, we find $\beta_{\rm{vir}}$ to be 0.05 - 0.79 (average $\sim$ 0.25).
For DVSOs, the $\beta_{\rm{rot}}$ values range from 0.33 to 0.79, which are comparable to the result obtained by Takahashi et al. (2006), 
who found a core with a large velocity gradient in the OMC-3 region.
The mean $\beta_{\rm{rot}}$ for the DVSOs is a factor of 3.7 higher than the mean of the non-DVSO sources, not a factor of 10.
We also estimated $\beta_{\rm{vir}}$ and the derived values are listed in Table \ref{beta}.
The parameter $\beta_{\rm{vir}}$ is an indicator of gravitational stability; $\beta_{\rm{vir}}$=0 for 
gravitationally bound state, $\beta_{\rm{vir}}$$<$0 for contracting, and $\beta_{\rm{vir}}$$>$0 for being unbound (expanding).
We found the obtained $\beta_{\rm{vir}}$ values range from $-$0.44 to 1.9 (average $\sim$ 0.39).
The objects with large $\beta_{\rm{rot}}$ values tend to have positive $\beta_{\rm{vir}}$ values, 
which result from the contribution of rotational energy to a gravitationally unbound condition, 
even if we take account of uncertainty of $\beta_{\rm{vir}}$.
Fifty percent of our targets have $\beta_{\rm{vir}}$ that are less than or close to zero, which implies that they are gravitationally bound. 
For NGC 7129 and Serpens, which are classified into $\it{Type \ C}$, the derived $\beta_{\rm{rot}}$ values are positive.
This result suggests that the gas dispersal of the clump leads the gravitationally unbound conditions.

\subsection{Clump-clump collision as a possible triggering of cluster formation}\label{4-3}

As discussed in the previous section, we found the following kinematic signatures in four out of the 13 $\HCO$ clumps (DVSOs);
(1) there exist distinct velocity gradients, 
(2) they have multiple components with different velocities on the rigid-like rotation, and 
(3) the clusters are located at the overlapped areas of the blue-shifted and red-shifted components in Figure \ref{maprb}. 
Thus, we regard the above kinematic condition as being one of the key structures in cluster formation, 
especially for the clusters including massive stars. We note the characteristics of DVSOs as follows.
The velocity gradients are seen in the objects for all three morphological types. 
Although whether the velocity gradient of clumps with rotational motion is observable depends on 
the orientation of the rotational axis to the line-of-sight, the uniformly distributed detection in the $\it{Types}$ 
of objects seems to indicate less correlation of these velocity structures with the three morphological types. 
We assume the velocity gradients as the rotational motions of the dense gas in the clumps, 
the contribution is a large portion of the internal energy within the clumps.
As a result, the clumps tend to be in a gravitationally unbound state. 
Such cluster formation occurring in a gravitational unbound dense clump is somewhat puzzling.
These observational results imply that these velocity structures are related to the initial conditions of cluster formation 
in the dense clumps, and must be preserved during the ages ($\sim$ 1--3 Myr) of our targets. 
We propose a clump-clump collision model as a possible mechanism for triggering formation of clusters to reproduce our observational results on the $\HCO$ clumps.

One possibility that can solve this puzzle is a clump-clump collision. 
We interpret the obtained data as the post-interacted (collision) stages of the clumps. 
Clump-clump collision can be responsible for triggering cluster formation through gas compression at the overlapped areas 
of the collision; shock heating at the overlapped areas can increase Jeans mass $(\propto {T}^{3/2}/\sqrt{\rho})$ and lead to form high-mass stars. 
The offset collision which is more common than head-on is also responsible for producing the angular momentum in 
merged clumps. Such a merged clump system is not necessary to be gravitationally bound for inducing cluster formation, 
although the collision velocity should not be too high to destroy the clump system. 
If the clump-clump collision induces the angular momentum of the merged clump system, the dynamical timescale of the rotation, 
$\tau_{\rm{rot}}$ is estimated as $\tau_{\rm{rot}}$=2$\pi/\omega$, where $\omega$ is the angular velocity of the clump.
For the typical angular velocity of DVSOs, $\omega$$\sim$2$\times$10$^{-13}$ s$^{-1}$, 
the timescale, $\tau_{\rm{rot}}$ is calculated to be $\sim$ 1$\times$10$^{6}$$\,$yr, which is comparable to the free-fall time.
If the rotational motion survives for the same period as the dynamical time scale, 
then the gradient can be observed for $\it{Type \ C}$ clumps, which have ages of $\sim$ 1--3 Myr \cite{lad03}.
The remnant of the collided clumps would also survive in the merged clump system and would be observed as multiple velocity components in $\HCO$.

Recently, dynamical interaction of massive clouds to induce cluster formation has been proposed in observational 
and theoretical studies. In the previous results, physical features similar to our results have been found.
Furukawa et al. (2009) proposed a scenario of cloud-cloud collision toward the cluster of Westerlund 2. 
They found that the Westerlund 2 is located between the two CO($J$=2--1) clouds, although their sizes and 
masses are an order of magnitude larger than those of our observed clumps. 
They found that the CO clouds are not gravitationally bound and expanding outward from the clusters, and proposed that 
the dynamical interaction between the clouds which promoted the gravitational instabilities form the Westerlund 2 cluster. 
For S87E, which is one of our observed targets, Xue $\&$ Wu (2008) suggested from the distributions in the multiple wavelength maps 
that the cluster formed by the clump-clump collision. 
Their results show that two clumps are approaching and the cluster is located at the overlapped area between the two clumps.
In addition, they suggest that the distribution of the compact H {\sc ii} region supports the above scenario.
While such observational evidence for cloud-cloud collision exists, 
the previous studies were mainly based on the spatial distribution between clouds and clusters in most cases.
In the theoretical perspective, Bhattal et al. (1998) simulated the phenomena of collisions of two clumps 
(size $\sim$1 pc, mass $\sim$100$\,$$\MO$, collision speed $\sim$1$\,$km$\,$${\rm{s}}^{-1}$) in a wide range of impact parameters. 
Their results show rigid-like rotations of the gas motion induced by clump-clump collision, 
for isothermal spheres, in the collision with a small impact parameter (the impact parameter is 0.3).
In their simulation, the clump-clump collisions produce multiple stars from B6 to B0 stars. 
The spatial and kinematic properties in our results are consistent with those of Bhattal et al. (1998).
Their calculated $\beta_{\rm{rot}}$ are 0.1--0.2 during the interaction.
We also estimated $\beta_{\rm{rot}}$ values of our observed objects adopting $p=2$,
and the derived $\beta_{\rm{rot}}$ values are consistent with the results of Bhattal et al. (1998). 
These consistencies indicate that cluster formation triggered by clump-clump collision can reasonably explain the four $\HCO$ clumps with velocity gradients, forming massive cluster members.

Another possibility is turbulent fragmentation.
The kinematic signatures are produced even in the early phase of clusters 
by the internal activities of the cluster members (e.g., outflows from newly born stars) or the external effects 
(e.g., nearby H {\sc ii} regions, supernovae etc.), which drive the turbulence energy toward dense clumps and enhance a turbulent fragmentation.
This mechanism may induce multiple cloud components with different velocities.
However, our targets are not contained in the catalogs of galactic supernovae \cite{gre09} and bubbles, 
which are detected in $\it{Spitzer}$/GLIMPSE survey \cite{chu06}, 
by contrast the Cassiopeia A \citep[e.g.,][]{mae09} or W3/W4/W5 regions \citep[e.g.,][]{all05, niw09}.
There are no spatial correlations between the areas of large velocity dispersion and the outflow lobes 
(see the right panel of Figure \ref{maph1} to \ref{maph5}). 
It takes a long time to produce large velocity gradients and these effects are hardly detectable unless they appear in evolved phases of 
cluster formation including an energetic internal source.
This is because the efficiency to convert the outflow energy into the turbulence energy of the dense clumps is generally expected to be low 
(a few $\%$ or less; Dyson $\&$ Williams 1997).
Therefore, it is difficult to explain that the relatively young clumps categorized in $\it{Type \ A}$ (e.g., AFGL 5142) 
have the distinct velocity gradient.

The cloud-cloud (clump-clump) collision mechanism is previously considered as an efficient mechanism to trigger star formation, 
but it also leads the destruction of the cluster-forming systems rather than gravitational collapse \citep{gil84}. 
The fraction of star formation triggered by cloud-cloud collision may be low in our Galaxy \citep{elm98}.
However, our results show that there are many more possibilities, especially to form massive stars 
by the clump-clump collision mechanism, which might occur inside the giant molecular clouds.
In previous studies, no clear velocity gradients have been detected and the velocity structure has not 
been discussed for the cluster-forming clumps. 
Our results suggest that it is important to observe more dense regions in the cluster-forming clumps and 
investigate the velocity structures to understand the formation mechanism of clusters.
In addition, to understand the initial conditions of the cluster formation, we would need to focus on the objects which are 
considered to be in younger phases of cluster formation (e.g., High mass protostellar objects, and Infrared dark clouds).

\section{CONCLUSION}

We have carried out a survey toward the 14 embedded clusters in the $\HCO$($J$=1--0) line with the Nobeyama 45m radio telescope. 
The purpose is to determine the evolutionary stages of the cluster-forming clumps and to understand 
the formation mechanism of massive clusters from the obtained physical properties. 
Our results and conclusions are summarized as follows.

\begin{enumerate}

\item We made the $\HCO$ maps with a size of $\sim$ 5$^{\prime}$ $\times$ 5$^{\prime}$--10$^{\prime} \times 10^{\prime}$ 
of the 14 nearby ($D$ $\leq$ 2.1$\,$kpc) cluster-forming regions. 
As a result, 13 clusters out of our targets are found to be associated with the $\HCO$ clumps, 
whose radii, masses, and velocity widths are 0.24--0.75$\,$pc, 100--1400$\,$$\MO$, and 1.5--4.0$\,$km s$^{-1}$, respectively.

\item We classified the $\HCO$ clumps into three types ($\it{Type \ A}$, $\it{Type \ B}$, and $\it{Type \ C}$) 
based on the spatial relation between the structures of $\HCO$ clumps and the locations of clusters using the same method in Higuchi et al. (2009).
The evolutionary stages determined from the $\HCO$ data are consistent with those from the $\CO$ data,
which implies that our classification scheme is robust.

\item
The tendency of SFEs estimated from the $\HCO$ data to be larger along the types from $\it{A}$ to $\it{C}$ 
is similar to that from the $\CO$ data.
The discrimination of SFE distribution among the evolutionary types is more conspicuous in $\HCO$ data than in $\CO$ data. 
This result indicates that $\HCO$ traces the dense regions in the $\CO$ clumps and physical conditions 
of dense regions are sensitive and affected by the clump-cluster evolution.

\item Four out of 13 $\HCO$ clumps, we named DVSOs have the distinct velocity gradients within the 1st moment maps. 
We found that the embedded clusters are located at the center of the velocity fields of dense clumps.
We estimated the virial ratios taking into account the contribution of rotation, 
and found that DVSOs are likely to be gravitationally unbound conditions.

\item In order to explain the physical conditions of DVSOs, we propose a clump-clump collision model 
as a possible mechanism for triggering cluster formation.

\end{enumerate}

\acknowledgments

We thank the referee for the constructive comments that have helped to improve this manuscript.
We acknowledge Daisuke Iono for his contribution to our study.
We thank Yoshimi Kitamura and Kazuyoshi Sunada for useful discussion about our data.
We are grateful to the staff of the Nobeyama Radio Observatory 
(NRO)$\footnote{Nobeyama Radio Observatory is a branch of the National Astronomical Observatory of Japan, National Institutes of Natural Sciences.
}$ for both operating the Nobeyama 45m telescope and helping us with the data reduction.
In particular, we acknowledge Nario Kuno, Shuro Takano, Tsuyoshi Sawada, and Takashi Tsukagoshi for their contributions to our observations.

\clearpage

\clearpage

\clearpage

\clearpage

\clearpage

\clearpage

\clearpage

\clearpage

\clearpage

\clearpage

\clearpage

\clearpage

\begin{deluxetable}{l l l l l l l l l l c c c c c c c c c c c}
\tabletypesize{\scriptsize}
\rotate
\tablecaption{Physical parameters of the $\HCO$ clumps
\label{para}}
\tablewidth{0pt}
\tablehead{
\colhead{Source Name} & \colhead{RA (J2000)} & \colhead{Dec (J2000)} & {$D$ [pc]} & \colhead{$M_{\rm{cluster}}$ [$\MO$]} &
\colhead{$\RC$[pc]} & \colhead{$\dvC$[km $\mathrm{s}^{-1}$]} & \colhead{$N(\mathrm{H}_{2}$)[$\times$ 10$^{22}$ $\mathrm{cm}^{-2}$]} 
& \colhead{$\MCLU$[$\MO$]} & \colhead{$T_{\rm{ex}}$ [$K$]}}
\startdata
 S87E & 19:46:19.9 & 24:35:24 & 2100 & 180 & 0.75$\pm$0.4 & 4.0$\pm$0.4 & 4.0$\pm$0.2 & 1400$\pm$400 & 22\\
 S88B &  19:46:47.0 & 25:12:43 & 2000 & 120 & 0.37$\pm$0.2 & 2.7$\pm$0.4 & 4.8$\pm$0.3 &  420$\pm$120 & 32  \\
 Gem 4 &  06:08:41.0 & 21:30:49 & 1500 & 190 & 0.58$\pm$0.3 & 2.0$\pm$0.3 & 4.5$\pm$0.2 & 940$\pm$200 & 28  \\
 AFGL 5180 & 06:08:54.1 & 21:38:24 & 1500 & 60 & 0.43$\pm$0.2 & 2.4$\pm$0.3 & 4.3$\pm$0.2 & 480$\pm$100 & 24  \\
 GGD 12-15 & 06:10:50.9 & $-$06:11:54 & 830 & 73 & 0.44$\pm$0.1 & 2.0$\pm$0.3 &  2.9$\pm$0.2 & 340$\pm$50 & 19  \\
 S235AB  &  05:40:52.5 & 35:41:25 & 1800 & 220 & 0.61$\pm$0.3 & 1.7$\pm$0.3  & 2.3$\pm$0.1 & 520$\pm$130 & 24  \\
 AFGL 5142 &  05:30:45.6 & 33:47:51  & 1800 & 50 & 0.54$\pm$0.3 & 3.6$\pm$0.3 &  3.1$\pm$0.2 & 550$\pm$140 & 17  \\
 Mon R2 & 06:07:46.6 & $-$06:22:59 & 830 & 340 & 0.75$\pm$0.2 & 1.9$\pm$0.3 &  2.8$\pm$0.2 & 930$\pm$140 & 19 \\
 S140 & 22:19:18 & 63:18:48 & 900 & 12 & 0.42$\pm$0.1 & 2.0$\pm$0.3 &  6.3$\pm$0.4 & 690$\pm$100 & 24  \\
 AFGL 490 &  03:27:38.7 & 58:46:58 & 910 & 25 & 0.45$\pm$0.1 & 1.8$\pm$0.3 &  2.1$\pm$0.1 & 250$\pm$40 & 16 \\
 BD+40$^{\circ}$4124 & 18:29:56.8 & 01:14:46 & 900 & 12 & 0.30$\pm$0.1 & 1.5$\pm$0.1 &  1.9$\pm$0.1 & 100$\pm$20 & 15 $\tablenotemark{a}$ \\
 NGC 7129 & 21:43:02 & 66:06:29 & 1000 & 76 & 0.48$\pm$0.2 & 1.5$\pm$0.4 &  1.4$\pm$0.1 & 140$\pm$30 & 12  \\
 Serpens SVS2 & 18:29:56.8 & 01:14:46 & 260 & 27 & 0.24$\pm$0.1 & 1.7$\pm$0.3 & 3.0$\pm$0.2 & 110$\pm$10 & 14  \\

\hline

\enddata
{\small
\tablecomments{This table shows the name, center coordinates, distance, mass of the clusters (from Lada $\&$ Lada (2003), Porras et al. (2003)),
$\RC$: radius, $\dvC$: velocity width, $N(\mathrm{H}_{2})$: mean $\mathrm{H}_{2}$ column density, 
$\MCLU$: clump mass,  $T_{\rm{ex}}$: excitation temperature derived by kinetic temperature of NH$_{3}$ of the clumps.}}
\tablenotetext{a}{Because we could not estimate the excitation temperature for low signal-to-noise ratio of the NH$_{3}$ data,
we adopted 15$\,\rm K$ which is the typical temperature in intermediate regions.}
\end{deluxetable}

\clearpage

\begin{deluxetable}{l l l l l l l l l l l c c c c c c c c c c c c}
\tabletypesize{\scriptsize}
\rotate
\tablecaption{Score sheet of the morphological classification of $\HCO$ clumps
\label{class}}
\tablewidth{0pt}
\tablehead{
\multicolumn{0}{c}{Source Name} & \colhead{Type} & \colhead{SFE} & \colhead{The Number of} & \colhead{Brightness} & \colhead{Offset Between} 
& \colhead{$\lambda_{\rm{J}}$} & \colhead{$\dvC(\HCO)$} & \colhead{$\MCLU(\HCO)$}  \\
\colhead{} & \colhead{} & \colhead{[$\%$]} & \colhead{Sub-clumps}  & \colhead{Contrast}  & \colhead{Clumps and Clusters [pc]} & \colhead{[pc]} & \colhead{$/\dvC(\CO)$} & \colhead{$/\MCLU(\CO)$}}

\startdata
 S87E & B & 12$\pm$3  & 2 & High & Off-peak : 0.3--0.6 & 0.17 & 1.2$\pm$0.1 & 0.85$\pm$0.05 \\
 S88B & C & 22$\pm$5  & 1 & Low & Off-peak : 0.5 $\&$ Cavity : $\sim$ 0.5 & 0.13 & 0.8$\pm$0.1 & 0.09$\pm$0.01 \\
 Gem 4 & B & 17$\pm$3 & 2 & High & Peak $\&$ Off-peak : 0.9 & 0.16 & 1.1$\pm$0.1 & 0.75$\pm$0.04 \\
 AFGL 5180 & B & 11$\pm$2  & 1 & High & Off-peak : 0.2 & 0.13 & 1.0$\pm$0.1 & 1.0$\pm$0.26 \\
 GGD 12-15 & B & 18$\pm$3  & $>$ 2 & Low & Off-peak : 0.3--0.5  & 0.14 & 0.9$\pm$0.1 & 0.25$\pm$0.01 \\
 S235AB & C & 30$\pm$6 & $>$ 2 & Intermediate & Off-peak : 0.3--0.5 & 0.21 & 0.8$\pm$0.1 & 0.40$\pm$0.07 \\
 AFGL 5142 & A & 8$\pm$2 &  1 & High & Peak  & 0.14 & 1.9$\pm$0.1 & 0.58$\pm$0.03 \\
 Mon R2 & C & 27$\pm$3 &  $>$ 5 & Low & Off-peak : 0.2--0.5 $\&$ Cavity : $\sim$ 0.3 & 0.19 & 0.9$\pm$0.1 & 0.77$\pm$0.07 \\
 S140 & A & 3$\pm$1 &  1 & High  & Peak & 0.11 & 0.8$\pm$0.1 & 0.37$\pm$0.01 \\
 AFGL 490 & B & 9$\pm$1 &  3 & Intermediate & Peak $\&$ Off-peak : 1--1.2 & 0.16 & 0.8$\pm$0.1 & 0.13$\pm$0.01  \\
 BD+40$^{\circ}$4124  & B & 10$\pm$2 &  2 & Low & Peak $\&$ Off-peak : 0.1--1 & 0.13 & 1.1$\pm$0.2 & 0.49$\pm$0.03  \\
 NGC 7129 & C & 35$\pm$5  & $>$ 4 & Low & Off-peak : 0.2--1.5 $\&$ Cavity : $\sim$ 0.5 & 0.20 & 1.0$\pm$0.2 & 0.19$\pm$0.01  \\
 Serpens SVS2 & C & 20$\pm$1 & $>$ 4 & High & Off-peak : 0.2--0.5 & 0.09 & 0.9$\pm$0.2 & 0.71$\pm$0.04  \\

\hline
\enddata
{\small
\tablecomments{
This table shows the source name, the classified type of the clump, the SFE derived by using the $\HCO$ mass, 
the number of sub-clumps, the brightness contrast of the clump, the location of the cluster, the Jeans length, 
the velocity width ratios of $\HCO$ to $\CO$, and the mass ratios of $\HCO$ to $\CO$.
The location of the cluster is written as follows, Peak: The emission peak of the clump which is associated with the cluster;
Off-Peak: The emission peak of the clump which is not associated with the cluster. 
The offset scale between the peak and the cluster is subsequently-shown [pc]; 
Cavity: In case that the cluster is located in a cavity. 
The size of the cavity is subsequently-shown [pc]. }}
\end{deluxetable}

\clearpage

\begin{deluxetable}{l l l l l l l c c c c c c c c}
\tabletypesize{\scriptsize}
\rotate
\tablecaption{The virial parameters of the $\HCO$ clumps 
\label{beta}}
\tablewidth{0pt}
\tablehead{
\multicolumn{0}{c}{Source Name} & \colhead{$\beta_{\rm{rot}}$} & \colhead{$\beta_{\rm{turb}}$} & \colhead{$\beta_{\rm{therm}}$} & \colhead{$\beta_{\rm{vir}}$} & \colhead{$\sigma_{\rm{grad}}$} & \colhead{$M_{\rm{star-max}}$$\tablenotemark{a}$} & \colhead{References} \\
\colhead{} & \colhead{} & \colhead{} & \colhead{[$\times$10$^{-8}$]} & \colhead{} & \colhead{[km $\rm{s}^{-1}$ pc$^{-1}$]} & \colhead{[$\MO$]} & \colhead{}}

\startdata
 S87E & 0.79$\pm$0.25 & 0.53$\pm$0.16 & 2.4 & 1.6$\pm$0.51 & 4.3$\pm$0.9 & 20$\pm$4 & 1 \\
 S88B & 0.34$\pm$0.11 &  0.56$\pm$0.17 &  5.8 & 0.81$\pm$0.25 & 2.2$\pm$0.4 & 20$\pm$4 & 2 \\
 Gem 4 & 0.05$\pm$0.01 &  0.23$\pm$0.06 & 3.5  & $-$0.44$\pm$0.10 & 1.0$\pm$0.2 & 19$\pm$4 & 3 \\
 AFGL 5180 & 0.33$\pm$0.08 & 0.47$\pm$0.11  & 4.4 & 0.6$\pm$0.14 & 2.2$\pm$0.4 & 18$\pm$4 & 4 \\
 GGD 12-15 & 0.05$\pm$0.01 &  0.48$\pm$0.08 & 4.9 & 0.07$\pm$0.01 & 0.7$\pm$0.1 & 15$\pm$3 & 5 \\
 S235AB & 0.09$\pm$0.02  &  0.29$\pm$0.08 & 5.7 & $-$0.24$\pm$0.07 & 1.0$\pm$0.2 & 15$\pm$3 & 6 \\
 AFGL 5142 & 0.55$\pm$0.15 &  0.91$\pm$0.25 & 3.2 & 1.9$\pm$0.53 & 2.7$\pm$0.5 & 10$\pm$2 & 7 \\
 Mon R2 & 0.18$\pm$0.03 &  0.13$\pm$0.02 & 3.1 & $-$0.38$\pm$0.07 & 1.7$\pm$0.3 & 10$\pm$2 & 8 \\
 S140 & 0.05$\pm$0.01 & 0.22$\pm$0.04 &  3.0 & $-$0.46$\pm$0.08 & 1.0$\pm$0.2 &  10$\pm$2 & 9 \\
 AFGL 490 & 0.20$\pm$0.03 & 0.28$\pm$0.05 & 5.9 & $-$0.04$\pm$0.01 & 1.2$\pm$0.2 & 8$\pm$2 & 10 \\
 BD+40$^{\circ}$4124  & 0.05$\pm$0.01 & 0.55$\pm$0.11 & 8.7 & 0.22$\pm$0.04 & 0.5$\pm$0.1 & 8$\pm$2 & 11  \\
 NGC 7129 & 0.23$\pm$0.05  & 0.55$\pm$0.12 & 8.5 & 0.56$\pm$0.13 & 0.9$\pm$0.2 & 8$\pm$2 & 12 \\
 Serpens SVS2 & 0.33$\pm$0.03 & 0.60$\pm$0.05 & 6.5 & 0.85$\pm$0.08 & 1.4$\pm$0.3 & 3$\pm$1 & 13 \\

\hline
\enddata
\tablerefs{
(1) Chen et al. (2004) ;
(2) Garay et al. (1993) ;
(3) Ghosh et al. (2000) ;
(4) Minier et al. (2005) ;
(5) Fang $\&$ Yao (2004) ;
(6) Felli et al. (1997) ;
(7) Zhang et al. (2002) ;
(8) Carpenter (2000) ;
(9) Preibisch \& Smith(2002) ;
(10) Schreyer et al. (2002) ;
(11) Looney et al.(2006) ;
(12) Fuente et al. (2002) ;
(13) Kaas et al. (2004)}
{\small
\tablecomments{This table shows that the the virial ratios of rotation, turbulent, thermal, and the combination of the above energies, 
which are derived adopting $p=0$, respectively, the velocity gradients within the clumps, the highest stellar mass of the cluster members, 
and the references of the highest stellar mass. }}
\tablenotetext{a}{The highest stellar mass of the cluster members.}
\end{deluxetable}

\end{document}